\documentclass{aastex7}
\usepackage[utf8]{inputenc}
\usepackage{placeins}
\usepackage{longtable}
\usepackage{amsmath}
\usepackage{subcaption}
\defcitealias{Erkal2019}{E19}

\begin{document}

\title{Estimate of Current Mass of the Large Magellanic Cloud from the Orphan-Chenab Tidal Stream}

\author{Hiroka T. Warren}
\affiliation{Department of Physics, Applied Physics and Astronomy, Rensselaer Polytechnic Institute, Troy, NY, 12180, USA}
\email{warreh2@rpi.edu}

\correspondingauthor{Hiroka T. Warren}
\email{warreh2@rpi.edu}

\author{Heidi Jo Newberg}
\affiliation{Department of Physics, Applied Physics and Astronomy, Rensselaer Polytechnic Institute, Troy, NY, 12180, USA}
\email{newbeh@rpi.edu}

\author{Autumn G. Guffey}
\affiliation{Department of Physics, Applied Physics and Astronomy, Rensselaer Polytechnic Institute, Troy, NY, 12180, USA}
\affiliation{Johns Hopkins University Applied Physics Laboratory, Laurel, MD, 20723, USA}
\email{autumnguffey44@gmail.com}

\author{JiaZhao Lin}
\affiliation{Department of Physics, Applied Physics and Astronomy, Rensselaer Polytechnic Institute, Troy, NY, 12180, USA}
\email{JiaZhao.Lin@rice.edu}

\author{Eric J. Mendelsohn}
\affiliation{Department of Physics, Applied Physics and Astronomy, Rensselaer Polytechnic Institute, Troy, NY, 12180, USA}
\affiliation{Department of Physics and Astronomy, Rutgers University, Piscataway, NJ, 08854, USA}
\email{rubegoldberg715@gmail.com}

\author{Kevin Roux}
\affiliation{Department of Physics, Applied Physics and Astronomy, Rensselaer Polytechnic Institute, Troy, NY, 12180, USA}
\email{rouxk@rpi.edu}

\begin{abstract}   

By fitting the tilt in the path of the Orphan-Chenab Stream (OCS), we conclude that the current mass of the Large Magellanic Cloud (LMC) within 30 kpc is $4.7$--$5.1 \times 10^{10}$ M$_\odot$. We note that the tidal radius of the LMC of this mass is 16.9 kpc, indicating that our measured mass approximates the current bound mass of the LMC. Previous measurements of the LMC mass based on fitting the observed path of the OCS through the Milky Way (MW) halo reported the {\it total} mass of the LMC. We show that because the closest approach of the LMC to the OCS, where the gravitational perturbation of the stream path is the highest, is about 20 kpc, the mass of the LMC outside of 30 kpc is not constrained and depends entirely on the assumed radial profile at large radius. Our best-fit total mass varies between $4.5 \times 10^{10}$ and $2.2 \times 10^{11}$ M$_\odot$ or more, depending on the presumed radial profile of the LMC. We also show that previous measurements of the mass of the LMC that used a particle-spray method to simulate the path of the OCS suffered from systematic error because they assumed that all particles were stripped from the dwarf galaxy at the tidal radius; N-body simulations show that particles are actually released from a range of distances from the center of mass of the OCS. In contrast, the choice of MW potential has little effect on the estimated LMC mass from the OCS.

\end{abstract}

\section*{} 

\twocolumngrid

\section{Introduction} \label{sec:intro}

The Large Magellanic Cloud (LMC) is the largest satellite galaxy of the Milky Way (MW). The gravitational effect of this satellite on our galaxy depends heavily on the LMC's total mass, including its dark matter content. However, the estimates of the total mass of the LMC have changed significantly in the past decade, increasing in mass by an order of magnitude to approximately $10^{11}$ M$_\odot$.

An LMC of this mass will significantly affect the dynamics of our MW galaxy. First, it will affect the evolution of tidal streams in the halo. An LMC with a mass of $10 \%$ of the MW will significantly affect the distance, angular path, and line-of-sight velocity of a stream orbit \citep{Law2010}. Moreover, the LMC will induce proper motions perpendicular to a stream \citep{Erkal2018}. Second, the LMC could induce the dynamical perturbations that change the shape of the outer part of the MW halo \citep{Vera2013}. Third, the center of mass of the MW can be displaced significantly by interaction with an LMC that is more massive than $5.0 \times 10^{10}$ M$_\odot$ \citep{Gomez2015}. Lastly, the LMC can cause disequilibrium in the disk \citep{Laporte2018} and halo \citep{Petersen2021}. The disequilibrium could affect the measurements of the mass of the MW from the velocity dispersion of halo stars under the assumption of an equilibrium system \citep{Magnus2022}.

Before 1960, the LMC was assumed to be $1.0 \times 10^{10}$ M$_\odot$ \citep{Burke1957,Kerr1957}. Then, \cite{Hunter1969} reported that it needed to be at least $2.0 \times 10^{10}$ M$_\odot$ in order to explain the observed warp in the Milky Way disk. Direct measurement of the LMC mass from the velocity dispersion of the LMC star clusters also showed that the LMC mass was larger than $1.5 \times 10^{10}$ M$_\odot$ \citep{Schommer1992}. The presumed LMC mass then remained fairly constant until recently.

In the last decade, estimates of the mass of the LMC have increased markedly. \cite{Besla2007} first argued that the LMC and Small Magellanic Cloud (SMC) could be on their first infall into our galaxy. This implies that the outer part of the dark matter halo might not have been stripped yet. \cite{Kallivayalil2013} suggested that the LMC might be more massive than $1.0 \times 10^{11}$ M$_\odot$, which would support the first infall scenario if the LMC and SMC have been bound together for several Gyr. A very massive LMC of $2.5 \times 10^{11}$ M$_\odot$ was then estimated from timing arguments \citep{Penarrubia2016} and to explain the warp in the MW from live N-body models of the LMC in the first infall scenario \citep{Laporte2018}. A mass of $1.24 \times 10^{11}$ M$_\odot$ is also estimated from the apparent satellites of the LMC \citep{Erkal2020}. Similarly, a mass range of $1.0$--$2.0 \times 10^{11}$ M$_\odot$ was derived by simultaneously fitting the mass of the LMC and MW to the distribution function of tracers (globular clusters and satellite galaxies) in the MW halo \citep{Magnus2022}. 

This new view of the LMC has also been supported by new observational data. For instance, \cite{Besla2012} found that the infall of an LMC and SMC interacting pair with a total mass of $2.0 \times 10^{11}$ M$_\odot$ could explain the observed properties of the Magellanic Stream. Likewise, \cite{Deason2015} claimed that the recent infall of a massive LMC ($\sim$10$^{11}$ M$_\odot$) would explain the large number of LMC satellites detected in the MW halo by the Dark Energy Survey (DES).

More recently, Erkal et al. (2019, hereafter \citetalias{Erkal2019}) estimated the mass of the LMC from its strong effect on the path of the Orphan-Chenab Stream (OCS), a stream of stars in the MW halo that formed from the tidal disruption of a dwarf galaxy (DG). This study found an LMC mass of $1.38 \times 10^{11}$ M$_\odot$ by running mass simulations using the modified Lagrange Cloud Stripping (mLCS) technique \citep{Gibbons2014}, also known as the particle-spray model, to generate model OCS paths to compare with the measured positions of stream stars \citepalias{Erkal2019}. The follow-up work by \cite{Koposov2023} suggested the LMC's total mass was $1.29 \times 10^{11}$ M$_\odot$ using improved data. The same technique was applied to other streams in order to estimate the mass of the LMC. Using the Sagittarius stream, \cite{Vasiliev2021} found a total LMC mass of approximately $1.3 \times 10^{11}$ M$_\odot$, and \cite{Shipp2021} found that it was $1.4$--$1.9 \times 10^{11}$ M$_\odot$ by simultaneously fitting five streams in the MW halo.

Dynamical measurement of the LMC mass using tidal streams requires assumptions, such as the LMC shape, the MW shape, and the LMC orbit. All of the assumptions used in modeling contribute to systematic errors in estimates of the LMC mass. Although a massive LMC is currently supported by many researchers, some studies suggest that a low mass LMC could also explain the observed properties of the Magellanic cloud \citep{Diaz2012, Wang2019, Wang2022}. Importantly, a low mass LMC is not ruled out.

Here we investigate the uncertainty and systematic errors in estimates of the LMC total mass from its strong effect on the path of the OCS (\citetalias{Erkal2019}), and conclude that while estimates of the current \textit{total} mass of the LMC are highly uncertain, the LMC mass within 30 kpc is well constrained to be within $4.7$--$5.1 \times 10^{10}$ M$_\odot$. 

In Section \ref{sec:simulation}, we show that particle-spray modeling does not accurately reproduce the path of the OCS stream. In Section \ref{sec:potential}, we show that the choice of reasonable MW potential models does not have much of an effect on estimates of the LMC mass from fits to the OCS. In Section \ref{sec:two_component}, we show that using an OCS progenitor DG with different radial profiles for the baryons and the dark matter has little effect on the path of the orbit and therefore the estimation of LMC mass. In Section \ref{sec:simulation_method}, we quantify how much systematic error in the LMC mass estimate is contributed by the use of the particle-spray method. In Section \ref{sec:LMC_model}, we show that the estimated LMC mass is heavily influenced by the radial profile of the LMC. In Section \ref{sec:LMC_mass}, we estimate the current LMC mass. In Section \ref{sec:discussion}, we discuss the implications of our findings on LMC mass estimation. The conclusions are presented in Section \ref{sec:conclusion}.

\section{Particle-spray vs N-body} \label{sec:simulation}

Here, we show that the simulation method (particle-spray or N-body simulation) can affect the observed paths of tidal streams and therefore the measurement of the mass of the LMC. In this section, we first create the OCS without the LMC by the particle-spray method, and then we compare it with an N-body simulation.

\begin{table}[]
    \centering
    \caption{Structural Models of the Galaxies.}
    \begin{tabular}{llr}
        \hline \hline 
        \textbf{Milky Way Potential} \\ \hline \hline
        \textit{Hernquist bulge:} & Mass: & 4.5$\times 10^9$ M$_\odot$ \\
        & Scale radius: & 0.442 kpc \\ 
        \hline
        \textit{Miyamoto-Nagai disk:} & Mass: & 6.8$\times 10^{10}$ M$_\odot$ \\
        & Scale length: & 3.0 kpc \\
        & Scale height: & 0.28 kpc \\
        \hline
        \textit{NFW halo:} & Mass: & 4.37$\times 10^{11}$ M$_\odot$ \\
        & Scale length: & 16.0 kpc \\ 
        \hline \hline 
        \textbf{DG Potential} \\ \hline \hline
        \textit{Plummer sphere:} & Mass: & $10^7$ M$_\odot$ \\
        & Scale radius: & 1.0 kpc \\ 
        \hline \hline 
        \multicolumn{3}{l}{\textbf{DG Present-Day Coordinates}} \\ \hline \hline
        & $l$: & 299$^\circ$ \\
        & $b$: & 5.75$^\circ$ \\ 
        & $d$: & 17.8 kpc \\
        & $v_{x}$: & -231.6 km/s \\
        & $v_{y}$: & 5.3 km/s \\ 
        & $v_{z}$: & 227.4 km/s \\
        \hline
    \end{tabular}
    \label{tab:N-body}
\end{table} 

\begin{figure*}
    \centering
    \textbf{No LMC}\par\medskip
    \includegraphics[width=0.9\textwidth]{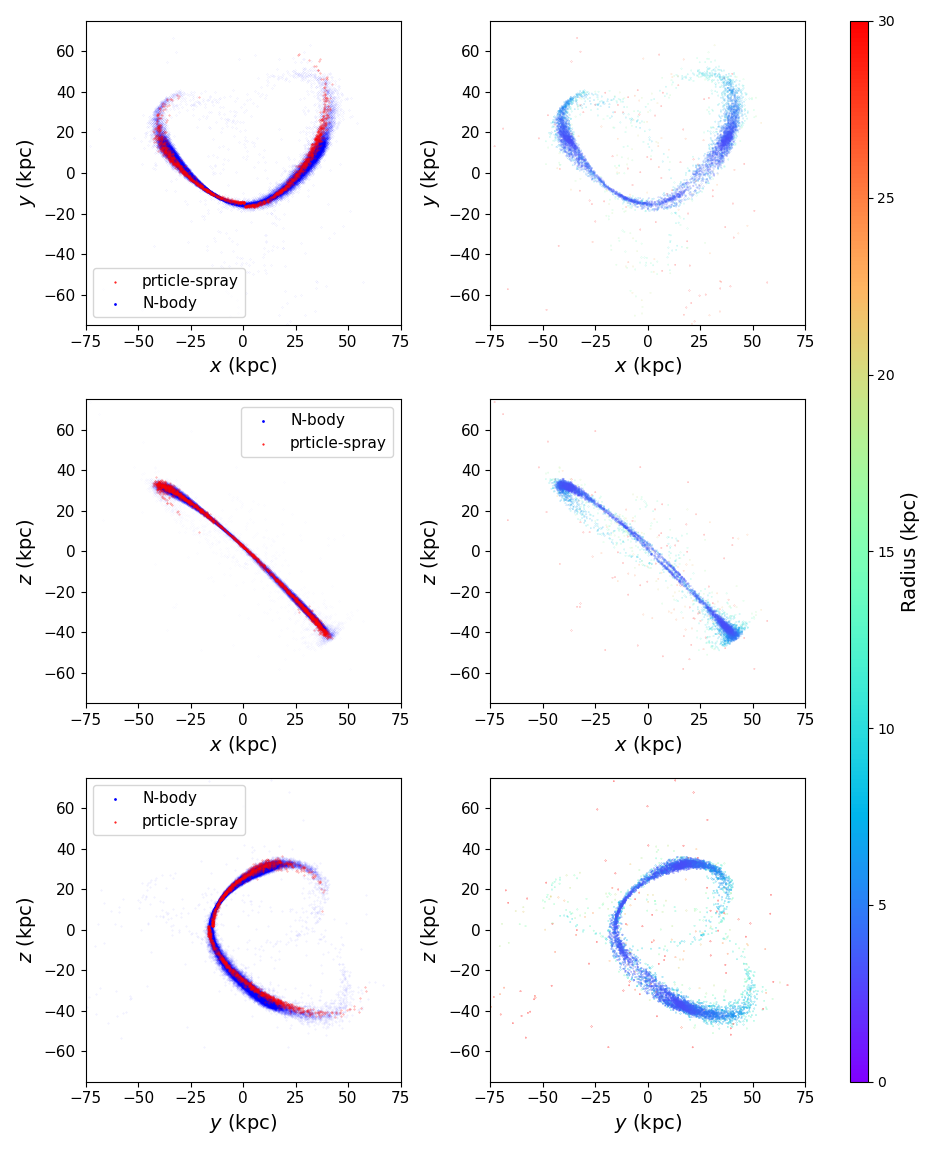}
    \caption{Left panels: Comparison of a simulated OCS without the LMC using N-body (blue) and particle-spray (red) methods using the same parameters (summarized in Table \ref{tab:N-body}). Right panels: Particles in the N-body simulation that stripped from a radius of greater than 3.0 kpc from the center of the DG. Note that the stream generated by the N-body simulation is wider, and the ends of the stream in particular follow a different path through space. Also, the N-body simulation stream is slightly tilted from the stream made with the particle-spray. The N-body particles that are stripped at a distance of more than 3 kpc from the DG center (right panels) exhibit the largest departure from the particle-spray simulation.}
    \label{Fig:OCS}
\end{figure*}

\subsection{Particle-spray Simulation} \label{subsec:particle-spray}

Using the particle-spray method, particles are released at the tidal radius as the progenitor of the stream moves in the gravitational potential. Therefore, there are essentially no particles inside of the progenitor. Since it does not need to calculate the internal motion of DG bodies, the particle-spray method is much more computationally efficient than the N-body simulation, which is highly time-consuming \citep{Gibbons2014}. The trade-off is that particle-spray is a less accurate approximation of the dynamical system, as will be shown in Section \ref{subsec:simulation_difference}. \citetalias{Erkal2019} used their own implementation of the particle-spray method to fit for the mass of the LMC.

To reproduce their results, we used the galpy package, \textit{streamspraydf} \citep{Fardal2015,Qian2022} for our particle-spray simulation of the OCS. Simulation parameters were selected to be similar to those used to model the OCS in \citetalias{Erkal2019}, with a small modification required because the galpy particle-spray (v1.80) does not have the LMC and the same MW potential options as \citetalias{Erkal2019}, and we want to use the same parameters for both particle-spray and N-body simulations so they can be directly compared. The mass and orbit parameters of the DG were selected from \citetalias{Erkal2019} (see Table \ref{tab:N-body}). Note that in particle-spray, there is no set scale radius because only the total mass of the DG and the MW potential are considered when determining the tidal radius at which particles are released from the simulated DG into the stream. 

For the MW potential, we chose \textit{MWPotential2014} from \cite{Bovy2015}, but we replaced the bulge profile with a Hernquist bulge \citep{Hernquist1990} with a mass of $4.5 \times 10^{9}$ M$_\odot$ and a scale radius of 0.442 kpc. The form of the bulge potential was modified because this potential is available in our N-body simulation code (which will be discussed in Section \ref{subsec:nbody}). This set of bulge parameters was chosen because \cite{Fardal2019} showed that this replacement has a negligible difference from \textit{MWPotential2014} beyond 10 kpc from the Galactic center. The stream was modeled with 1,000 particles, which were released at the tidal radius as the DG was evolved for 5.0 Gyr forwards in time. The effect of the LMC is not included in this simulation. The parameters are summarized in Table \ref{tab:N-body}, and the properties of the resulting stream are shown in red in the left panels of Figure \ref{Fig:OCS}. The implications of Figure \ref{Fig:OCS} will be discussed in Section \ref{subsec:simulation_difference}.

\subsection{N-body Simulation} \label{subsec:nbody}

We ran our N-body simulations with MilkyWay@home \citep{Shelton2018, Shelton2021, Mendelsohn2022}, which is publicly available on GitHub. MilkyWay@home N-body simulations use a Barnes-Hut tree algorithm \citep{Barnes1986} to approximate the Newtonian gravitational acceleration between particles. The threshold value is set to 1 \citep{barnes2001}. The softening length is calculated as:
\begin{equation}
\epsilon=\frac{R_{s}}{10\sqrt{N}},
\end{equation}
where $N$ is the number of bodies in a simulation, and $R_s$ is the scale length. For a two component model in Section \ref{sec:two_component}, we calculate a center of mass scale length as:
\begin{equation}
R_s=\frac{M_{b}a_{b}+M_{d}a_{d}}{M_{b}+M_{d}},
\end{equation}
where $M_b$ and $M_d$ are baryon and dark matter masses, and $a_b$ and $a_d$ are baryon and dark matter scale lengths \citep{Shelton2018}. The stability of the simulated DGs is explained in the Appendix.

The MW potential and the orbit parameters of the DG are the same as the particle-spray simulation (Table \ref{tab:N-body}). The DG was modeled by a single Plummer sphere \citep{Plummer1911} with a scale radius of 1.0 kpc, a progenitor mass of $10^{7}$ M$_\odot$ as in \citetalias{Erkal2019}, and 40,000 bodies \citep{Mendelsohn2022}. In the N-body simulation, the orbit parameters are used to place a single body at the starting point (present position). The body is then evolved for 5.0 Gyr backwards in time. At this position, it is replaced by the DG progenitor. We generate the DG progenitor using the technique described in the Appendix of \cite{Aarseth1974}. Then, the orbit integration is evolved forward until it matches the present stream position. Again, the effect of the LMC is not included in this simulation. The parameters are summarized in Table \ref{tab:N-body}, and the properties of the resulting stream are shown in blue in the left panels of Figure \ref{Fig:OCS}. 

\begin{figure}
    \centering
    \includegraphics[width=0.45\textwidth]{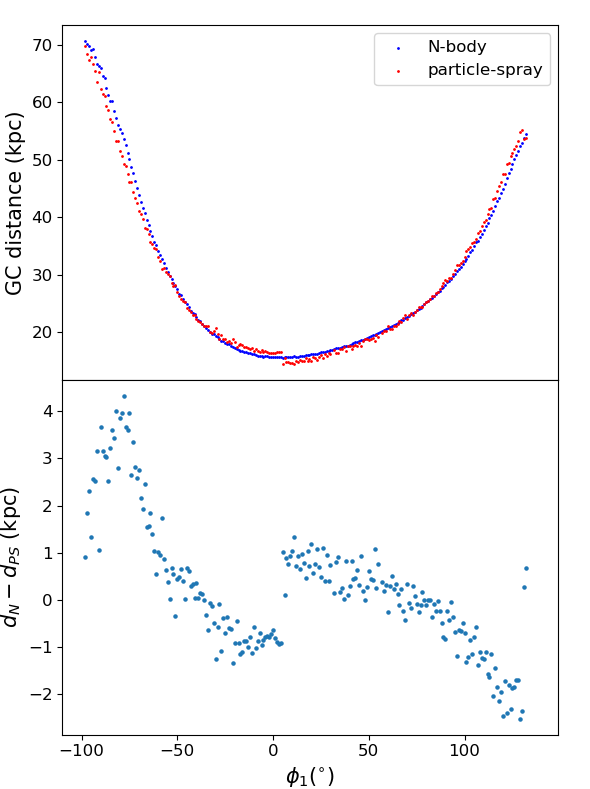}
    \caption{Deviation in the Galactocentric distance of the center of the particle-spray stream from the center of the N-body stream (no LMC). Note that the x-axis is Sun-centered along the stream. The upper panel shows that the particle-spray and N-body streams are tilted from each other. The lower panel shows the Galactocentric distance difference between the two simulations. This distance difference trends from the N-body simulation being farther at low $\phi_1$ to the particle-spray simulation being farther at large $\phi_1$, quantifing the stream tilt between the N-body and particle-spray simulations.}
    \label{Fig:tilt}
\end{figure}

\subsection{Comparison of OCS N-body and Particle-spray Simulations} \label{subsec:simulation_difference}

The left panels in Figure \ref{Fig:OCS} show the OCS without the LMC created by both the particle-spray technique and an N-body simulation, using the same parameters. The Galactic Cartesian coordinates $(x, y, z)$ are calculated in a right-handed frame where the Sun’s location is $(x, y ,z) = (-8.1, 0, 0)$ kpc. Although the results of the two simulations are similar to each other overall, note that the particle-spray stream is tilted from the stream created with the N-body simulation. The N-body simulation is much wider, and the ends of the streams in particular follow different paths through space. 

In Figure \ref{Fig:tilt}, we quantify how much the particle-spray stream is tilted from the N-body stream in simulations that do not include the LMC. We calculated the 3$\sigma$-clipped mean of particles in each stream for 1 degree bins in $\phi_{1}$. The $\phi_{1}$ (angle along the stream) and $\phi_{2}$ (angle perpendicular to the stream) sky coordinates for the OCS are defined by \cite{Koposov2019}. The top panel shows the Galactocentric distance of the resulting stream centers as a function of $\phi_{1}$. Note that the particle-spray method maintains an energy difference between the leading and trailing tidal tails (as evidenced by a discontinuity in the Galactocentric distance at $\phi_1$ = 0$^{\circ}$), which the N-body simulation does not. We then computed the difference in Galactocentric distance between the two simulated streams in each bin. The bottom panel of Figure \ref{Fig:tilt} shows the difference between the Galactocentric stream distances presented in the top panel, as a function of $\phi_1$. Note the negative trend in $d_{N}-d_{PS}$, which quantifies the stream tilt between simulations. In this analysis, the LMC is not taken account into the simulations. However, since the LMC mass estimate fitting a stream heavily depends on the fit of the simulated stream path to the data, the incorrect tilt in the particle-spray simulation would cause systematic errors in the mass estimate.

\begin{figure}
    \centering
    \includegraphics[width=0.45\textwidth]{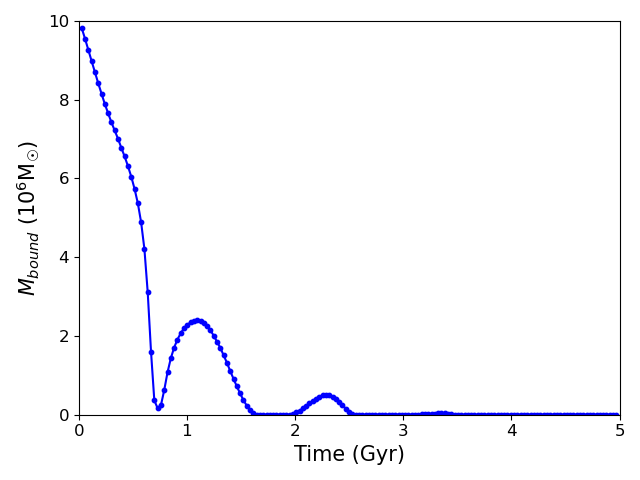}
    \caption{Bound mass of the DG in the N-body simulation (no LMC) as a function of time. Most particles become unbound at perigalacticon, and then some particles become bound again at larger distances from the Galactic center.}
    \label{Fig:bound_mass}
\end{figure}

\begin{figure}
    \centering
    \includegraphics[width=0.45\textwidth]{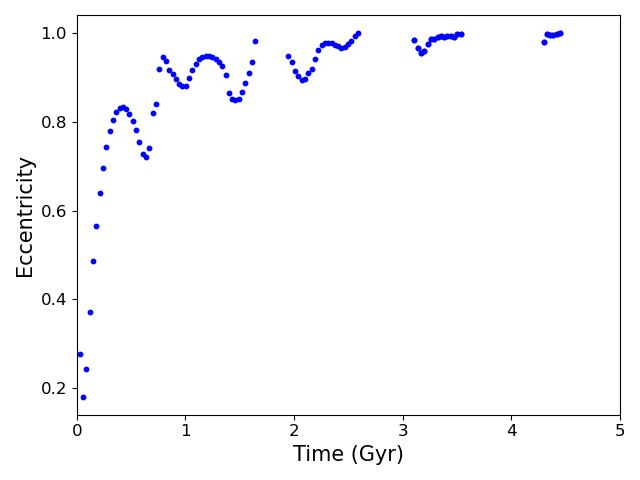}
    \caption{Eccentricity of the DG in the N-body simulation (no LMC) as a function of time. The DG shape quickly deviates from spherical, in contrast to the particle-spray which assumes a spherical DG at all times.}
    \label{Fig:DG_shape}
\end{figure}

To explore the cause of the stream tilt, we compared DG disruption and stream formation between the two models. We first identified the bound particles and calculated the bound mass for each time step of our N-body simulation. For a set of $N$ particles in a DG, the kinetic energy of the $i$-th particle is:

\begin{equation}
\mathit{KE}_{i} = \frac{1}{2} |\vec{v}_{i} - \vec{v}_{com}|^{2},
\end{equation}
where $|\vec{v}_{i} - \vec{v}_{com}|^{2} = (v_{i,x} - v_{com,x})^{2} + (v_{i,y} - v_{com,y})^{2} + (v_{i,z} - v_{com,z})^{2}$ and $\vec{v}_{com}$ is the center of mass velocity. Similarly, the potential energy between particles $i$ and $j$ $(i \neq j)$ is:

\begin{equation}
\mathit{PE}_{i} = - \sum_{j} \frac{m_{i}}{|\vec{r}_{i} - \vec{r}_{j}|} ,
\end{equation}
where $|\vec{r}_{i} - \vec{r}_{j}| = \sqrt{(x_{i} - x_{j})^{2} + (y_{i} - y_{j})^{2} + (z_{i} - z_{j})^{2}}$ and $m_{i}$ is the mass of each particle. The total energy of the $i$-th particle is given by:

\begin{equation}
\mathit{E}_{i} = \mathit{KE}_{i} + \mathit{PE}_{i} .
\end{equation}

If $E_{i} < 0$, then the particle is bound. We calculated the bound mass of the DG in each time step by adding up the mass of the bound particles at that time. Figure \ref{Fig:bound_mass} shows the bound mass as a function of time in the N-body simulation (no LMC). The bound mass of the DG quickly decreases around perigalacticon, but some particles become bound again as the DG moves farther away from the Galactic center and the tidal radius increases. This cycle continues until complete disruption at 4.45 Gyr.

We calculated the eccentricity of the DG in the N-body simulation. The eccentricity is:

\begin{equation}
e \tbond \sqrt{1 - \left( \frac{R_{min}}{R_{max}}\right)^{2}} ,
\end{equation}
where we defined $R_{min}$ and $R_{max}$ as the minimum and maximum axis lengths in a principle component fit of the bound particles in the DG. Figure \ref{Fig:DG_shape} shows the eccentricity (zero is a spherical DG) of the DG in each time step (no LMC). The DG quickly becomes very eccentric in shape in the N-body simulation, unlike the particle-spray technique which assumes a spherical DG at all times. 

\begin{figure}
    \centering   \includegraphics[width=0.4\textwidth]{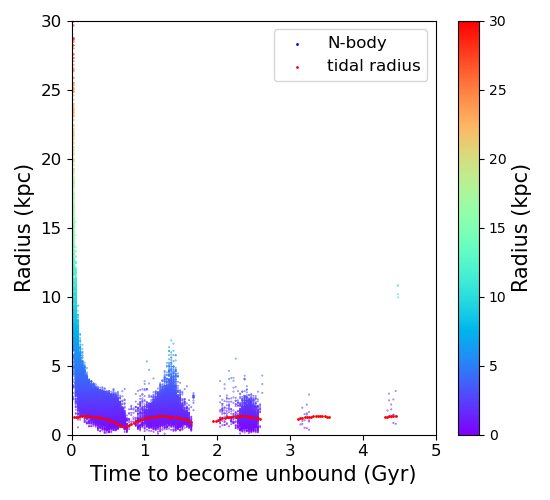}
    \caption{Distance from the center of mass of the DG at which particles become unbound for the last time as a function of time (no LMC). The red line shows the tidal radius with respect to time, where the particles are released using the particle-spray. Note that the N-body simulation produces a wide variety of radii while the particle-spray method by construction assumes only one radius at a time.}
    \label{Fig:disruption_radius}
\end{figure}

\begin{figure*}
    \centering
    \textbf{MW + LMC}\par\medskip
    \includegraphics[width=\textwidth]{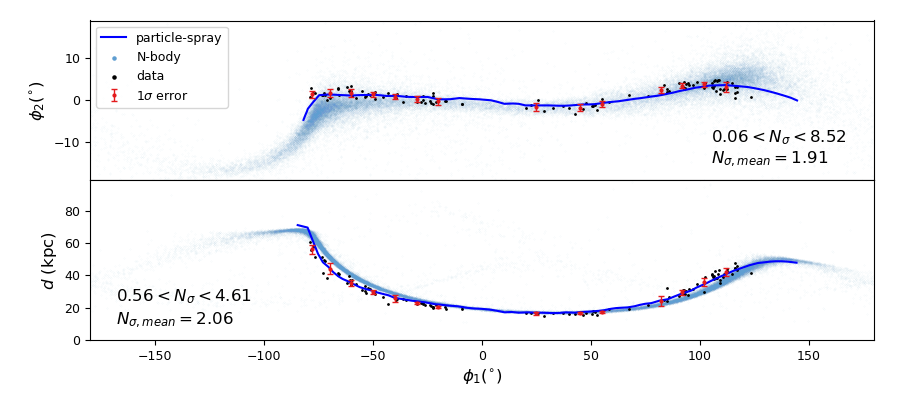}
    \caption{Comparison of the OCS simulated by particle-spray (blue line) and N-body simulation (light blue dots) with OCS data (black dots). The particle-spray line is from \citetalias{Erkal2019}. The N-body points were generated from a simulation with the same MW, orbit, and DG parameters as were used in the particle-spray calculation, including an LMC of mass of 1.2501$\times 10^{11}$ M$_\odot$. Black points are data from \cite{Koposov2019}. $N_{\sigma}$ is the number of $\sigma$ from the center of the stream (data) to the center of the simulated stream (N-body). The range and mean of deviations are given in the bottom right corner for the $\phi_2$ vs. $\phi_1$ panel and the bottom left corner for the $d$ vs. $\phi_1$ panel. The error bars (red) show the 1$\sigma$ errors we used to compute $N_{\sigma}$. The OCS N-body simulation deviates significantly from the particle-spray simulation. To make the N-body simulation match the data, it is necessary to change the assumed LMC mass. In Section \ref{sec:simulation_method}, we quantify how much the simulation technique itself influences the measurement of the mass of the LMC.}
    \label{Fig:nbody_particle_spray}
\end{figure*}

To investigate where the particles are released in the particle-spray modeling, we calculated the tidal radius. The tidal radius, $r_{t}$, that is used in the particle-spray simulation is \citep{Fardal2015}:

\begin{equation}
r_{t} = \left(\frac{GM_{DG}}{\Omega^{2} - \frac{\partial^{2}\Phi_{MW}}{\partial r^{2}}}\right)^{\frac{1}{3}} ,
\label{Eq:tidal_radius}
\end{equation}
where $\Omega = \left( \frac{GM_{MW}(<r)}{r^{3}}\right)^\frac{1}{2}$. $M_{DG}$ is the mass of the DG, $\Phi_{MW}$ is the MW potential. Here, $r$ is the distance from the Galactic Center to the center of mass of the bound particles in the DG, and $M_{MW} (<r)$ is the enclosed mass of the MW. 

Figure \ref{Fig:disruption_radius} shows the radii at which particles become unbound in both simulations (no LMC). For particle-spray, the tidal radius is shown as a function of time. For the N-body simulation, we tracked when each particle becomes unbound for the last time and obtained its distance from the center of mass. In contrast to the particle-spray which has one disruption radius, the N-body simulation shows a wide range of radii at which particles become unbound. Around perigalacticon, where most of the disruption occurs, the particles are stripped from a wide variety of radii, meaning that they strip with a wide variety of total energies. This causes the stream to become wider. The leading arm of the stream is closer to the Galactic center than the trailing arm of the stream because the particles in the leading arm have lower energy while the particles in the trailing arm have higher energy. Having a wider range of energies means the center of the tidal stream is pulled even closer (leading tail) or further (trailing tail) from the Galactic center. This causes the N-body simulation stream to be tilted from the generated particle-spray stream.

The right panels in Figure \ref{Fig:OCS} show the particles that become unbound at a radius greater than 3.0 kpc, which is much larger than the particle-spray tidal radius at any time. Figure \ref{Fig:OCS} shows that they would be expected to differ the most from the particle-spray simulation. These particles indeed mostly form the edges of both arms, making the stream wider in the N-body simulation. Note that the largest discrepancies are at the very ends of the stream.

We can now understand that the jump at $\phi_1\backsim0^\circ$ in Figure \ref{Fig:tilt} (and also visible in the lower left panel of Figure 1 near $z=0$) arises from the larger discontinuity in energy between the leading and trailing portions of the particle-spray simulation. This discontinuity stems from the stream particles being released from the highest and lowest potential energy points of a spherical approximation to the tidal radius; the N-body particles are stripped at a continuous range of energies from a highly elongated DG.

\subsection{Comparison of Simulated OCS Streams With Real Data} \label{sec:sigma_error}

Since the LMC mass measurement that is derived from the OCS depends heavily on its path, it is not surprising that the tilt between the simulation techniques in Section \ref{subsec:simulation_difference} would influence the LMC mass measurement. Figure \ref{Fig:nbody_particle_spray} shows a comparison between the N-body simulation and the particle-spray simulation, using the same LMC mass in both simulations. The particle-spray stream (blue line) is from \citetalias{Erkal2019} with a spherical NFW potential \citep{Navarro1997}. We ran an N-body simulation (light blue dots) with the same MW potential, DG orbit, progenitor properties, and LMC mass (1.2501$\times 10^{11}$ M$_\odot$) as \citetalias{Erkal2019}. The data showing the position of the OCS in the sky is also from \cite{Koposov2019}.

As one might expect based on our previous analysis, the simulated streams are not an exact match to each other in distance or position on the sky even though the same simulation parameters were used. While the stream made with particle-spray fits the data perfectly, the N-body simulation deviates from the data. In Section \ref{sec:simulation_method}, we show the N-body simulation can be made to fit the data by lowering the LMC mass, and we quantify how much the simulation technique itself influences the mass of the LMC measurement. As we will show, the estimated total LMC mass also depends on the assumed DG radial profile.

Throughout this paper, we show how well the simulated OCS fits to data by calculating the number of sigma, $N_\sigma$, from the center of data to the center of the simulated data. In order to do so, we divided the simulated OCS into bins of width 1$^\circ$ in $\phi_1$. For each bin, we calculated the 3$\sigma$-clipped mean value of $\phi_2$, the heliocentric distance ($d$), and the proper motions in RA ($\mu_{\alpha}$) and Dec ($\mu_{\delta}$). For the data, we first fit polynomials to the center of the real data points from \cite{Koposov2019}. In $\phi_{2}$, $\mu_{\alpha}$, and $\mu_{\delta}$, we fit two polynomials (the negative $\phi_1$ and the positive $\phi_1$), because of the complicated stream shape. In $\phi_{2}$, we fit 3rd ($< 0^{\circ}$) and 4th ($> 0^{\circ}$) order polynomials. In $\mu_{\alpha}$, we fit 7th ($< 0^{\circ}$) and 3rd ($> 0^{\circ}$) order polynomials. In $\mu_{\delta}$, we fit 6th ($< 0^{\circ}$) and 3rd ($> 0^{\circ}$) order polynomials. In $d$, we fit a 2nd-order polynomial to the whole data range.

We then selected a point every 10$^{\circ}$ in $\phi_{1}$ from the fits. We made $\pm$ 5$^{\circ}$ bins for these selected points and stored the real data points in each bin. However, we removed data bins when there were 3 data points or fewer in a 10$^\circ$ bin in $\phi_1$. We needed to remove data bins after the fit because they were all needed for the polynomial fit go through the center of the real data points.

We then computed the difference between each actual data point in the bin and the polynomial fit. The 1$\sigma$ error (red error bars in Figure \ref{Fig:nbody_particle_spray}), which indicates the standard error at each 10$^\circ$ bin, is equal to the square root of the average of the squares of these values. The number of sigma, $N_\sigma$, by which each data bin deviates from the simulation was found by calculating how many standard deviations the center of the simulated data deviates from the center of data, at the center of each bin.

In Figure \ref{Fig:nbody_particle_spray}, the center of the N-body simulation of the OCS differs from the center of the data by 0.06 $< N_{\sigma_{\phi_2}} <$ 8.52 ($N_{\sigma, mean}$=1.91) and 0.01 $< N_{\sigma_{d}} <$ 4.61 ($N_{\sigma, mean}$=2.06) for all of the measured stream positions. The large deviations indicate the N-body simulation is not a good fit to the data even though it used the same simulation parameters as were used for the well-fit particle-spray simulation.

\section{Effect of Milky Way Potential} \label{sec:potential}

In this section, we explore the relationship between the assumed MW potential and the measurement of the LMC total mass. We fitted the OCS data to simulations produced in four MW potential models including the LMC to determine the sensitivity of the LMC mass estimates to the choice of MW potential.

\subsection{Milky Way Potential Selection}

In order to compare the measurement of the LMC mass with different MW potential models, we selected models from \citetalias{Erkal2019}, \cite{Salas2019}, and \cite{Mendelsohn2022}. We also chose \textit{MWPotential2014} from \cite{Bovy2015}, but replaced the bulge profile with the Hernquist profile as in Section \ref{sec:simulation}. We hereafter refer to these MW potential models as \textit{Erkal2019}, \textit{Salas2019}, \textit{Mendelsohn2022}, and \textit{Modified MWPotential2014} (\textit{MMWPotential2014}). The parameters for each model are summarized in Table \ref{tab:MW_potential}.

\begin{figure}
    \centering
    \caption{Rotation curves of \textit{Erkal2019} (top), \textit{Salas2019} (second from top), \textit{Mendelsohn2022} (second from bottom), and \textit{MMWPotential2014} (bottom). The rotational velocity of each potential model is compared with the measured values from \cite{Eilers2019}. Note that \textit{Mendelsohn2022} exhibits the largest departure from the observed rotational velocities.}
    \includegraphics[width=0.45\textwidth]{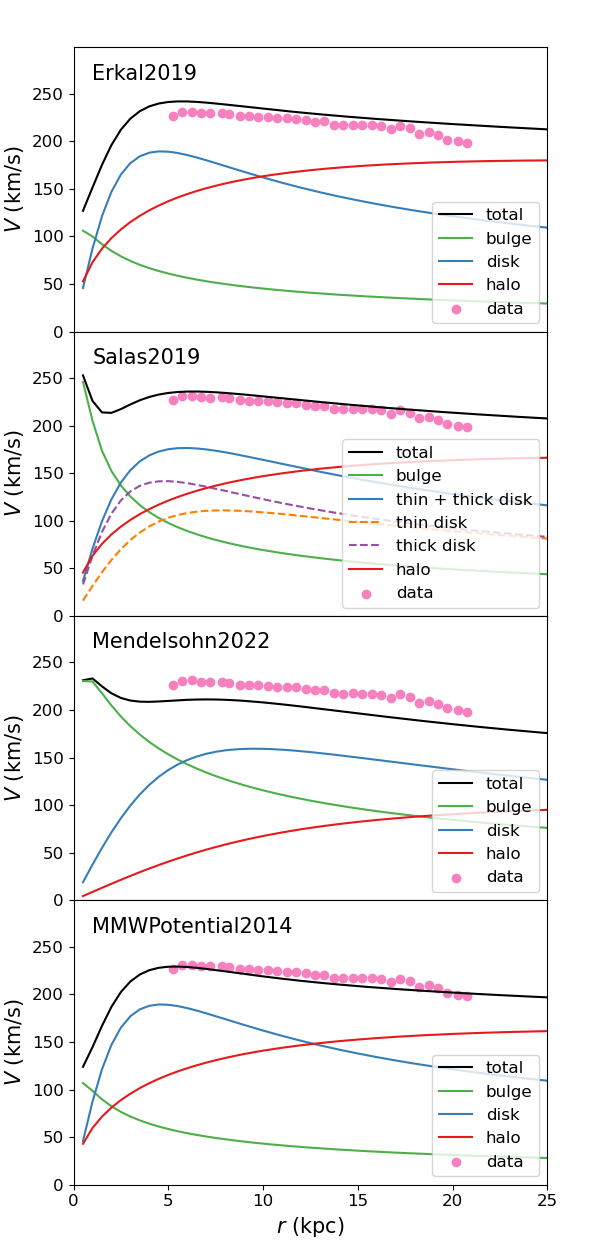}  
    \label{Fig:MW_potential}
\end{figure}

Figure \ref{Fig:MW_potential} shows the rotation curves of the MW in each model compared to the rotational velocity measurements of the MW from \cite{Eilers2019}. Note that they all roughly agree with the measured value, though \textit{Mendelsohn2022} is the most discrepant with a rotation curve that is lower than measured.

\begin{table}[]
    \centering
    \caption{Milky Way potential models and parameters used in Figure \ref{Fig:MW_potential}. The masses here are total masses.}
    \begin{tabular}{llr}
        \hline \hline 
        \textbf{Erkal2019} & & \\ \hline \hline
        \textit{Hernquist bulge:} & Mass: & 5.0$\times 10^9$ M$_\odot$ \\
        & Scale radius: &  0.5 kpc \\ \hline
        \textit{Miyamoto-Nagai disk:} & Mass: & 6.8$\times 10^{10}$ M$_\odot$ \\
        & Scale length: &  3.0 kpc \\
        & Scale height: &  0.28 kpc \\ \hline
        \textit{NFW spherical halo:} & Mass: & 7.916$\times 10^{11}$ M$_\odot$ \\
        & Scale length: &  12.807 kpc \\
        \hline \hline
        \textbf{Salas2019} & & \\ \hline \hline
        \textit{Plummer bulge:} & Mass: & 1.067$\times 10^{10}$ M$_\odot$\\
        & Scale radius: & 0.3 kpc \\ 
        \hline
        \textit{Miyamoto-Nagai thin disk:} & Mass: & 3.944$\times 10^{10}$ M$_\odot$ \\
        & Scale length: & 5.3 kpc \\
        & Scale height: & 0.25 kpc \\
        \textit{Miyamoto-Nagai thick disk:} & Mass: & 3.944$\times 10^{10}$ M$_\odot$\\
        & Scale length: & 2.6 kpc \\
        & Scale height: & 0.8 kpc \\ 
        \hline
        \textit{NFW halo:} & Mass: & 4.287$\times 10^{11}$ M$_\odot$\\
        & Scale length: & 14.9 kpc \\ 
        \hline \hline 
        \textbf{Mendelsohn2022} & & \\ \hline \hline
        \textit{Hernquist bulge:} & Mass: & 3.4$\times 10^{10}$ M$_\odot$\\
        & Scale radius: & 0.7 kpc \\ \hline
        \textit{Miyamoto-Nagai disk:} & Mass: & 9.911$\times 10^{10}$ M$_\odot$\\
        & Scale length: & 6.5 kpc \\
        & Scale height: & 0.26 kpc \\ \hline
        \textit{Logarithmic halo:} & V halo: & 73 km/s \\
        & Scale length: & 12.0 kpc \\
        \hline \hline 
        \textbf{MMWPotential2014} & & \\ \hline \hline
        \textit{Hernquist bulge:} & Mass: & 4.5$\times 10^9$ M$_\odot$ \\
        & Scale radius: & 0.442 kpc \\ 
        \hline
        \textit{Miyamoto-Nagai disk:} & Mass: & 6.80$\times 10^{10}$ M$_\odot$ \\
        & Scale length: & 3.0 kpc \\
        & Scale height: & 0.28 kpc \\ 
        \hline
        \textit{NFW halo:} & Mass: & 4.37$\times 10^{11}$ M$_\odot$ \\
        & Scale length: & 16.0 kpc \\ 
        \hline 
    \end{tabular}
    \label{tab:MW_potential}
\end{table} 

\subsection{Modeling the LMC in MilkyWay@home}

The LMC is modeled in our N-body simulations as a moving but otherwise rigid potential falling into a static MW potential, both represented via analytical equations. To determine the motion of the LMC and the MW's response, we take into account the effect of dynamical friction (DF) on the LMC, and the gravitational pull of the LMC on the MW itself. If the total LMC mass is $\sim$10$\%$ of the MW (Section \ref{sec:intro}), we expect a non-negligible gravitational effect on the center of mass of the MW. Due to the static implementation of the MW, we cannot move its position, and so we address this by instead applying an opposing shift to the LMC and the bodies that represent the DG at every time step. The MW centered coordinate system is thus a non-inertial reference frame. Our simulations consist of two stages: backwards integration from the present day starting position to determine a starting point for the simulation, and forwards integration from there to present day. 

We use an initial LMC position, in right-handed Galactic Cartesian coordinates of $(x, y ,z) = (-0.52, -40.8, -26.5)$ kpc \citep{Van2002,Pietrzyski2013}, and an initial velocity $(V_{x}, V_{y} ,V_{z}) = (-58.2, -231, 226)$ kpc/Gyr \citep{Kallivayalil2013}. The LMC can be modeled by either a spherical Plummer or a Hernquist profile, and its scale radius is selected so that the enclosed mass at 8.7 kpc agrees with the measured value of 1.7$\times 10^{10}$ M$_\odot$ \citep{Van2014}. The MW and LMC are rigid potentials in both stages of the simulation. As in Section \ref{subsec:nbody}, the DG is represented by a point mass during backwards integration, but it is then replaced with its N-body form at the start of forward integration. We retain the same initial orbit parameters for the DG.

We use a Velocity Verlet algorithm \citep{Verlet1967} to integrate bodies backwards in our simulation. We flip the signs of DG and LMC's velocities at their starting positions (present day) as we are integrating backwards. We then start by finding the accelerations experienced by the MW ($\textbf{a}_{0,MW}$), DG ($\textbf{a}_{0,DG}$), and LMC ($\textbf{a}_{0,LMC}$) at their current positions. We calculate the MW acceleration, $\textbf{a}_{0,MW}$, due to the LMC, from the selected density profile of the LMC (Plummer or Hernquist profile in this paper), that is:
\begin{equation}\label{eq:eq_lmc_plummer}
\textbf{a}_{P,t} = -\frac{GM_{LMC}}{(r^{2} + a^{2})^{\frac{3}{2}}}\textbf{r}_{t}
\end{equation}
or 
\begin{equation}\label{eq:eq_lmc_hern}
\textbf{a}_{H,t} = -\frac{GM_{LMC}}{r(r + a)^{2}}\textbf{r}_{t},
\end{equation}
where $M_{LMC}$ is the LMC total mass, $a$ is the scale radius, $\textbf{r}$ is the Galactocentric position vector at time step, $t$, and $r = |\textbf{r}_{t}|$. $\textbf{a}_{0,DG}$ is the acceleration due to the MW rigid potential \citep{Shelton2018, Shelton2021, Mendelsohn2022} plus either $\textbf{a}_{P,0}$ or $\textbf{a}_{H,0}$ from Eqn. \ref{eq:eq_lmc_plummer} and \ref{eq:eq_lmc_hern} but with  $\textbf{r}$ defined instead as $\textbf{r} = \textbf{r}_{LMC}-\textbf{r}_{DG}$. To calculate $\textbf{a}_{0,LMC}$, we need to include the DF on the LMC as well as the acceleration from the MW. We ignore the DG's gravitational impact on the LMC and MW because of its much smaller mass.  

We calculate the DF using \cite{Chandrasekhar1943}'s formula:
\begin{multline}
\textbf{F}_{DF} = -16\pi^{2} \ln(\Lambda)m_{s}(M_{LMC} + m_{s}) \frac{\textbf{v}_{LMC}}{v_{LMC}^{3}}\\
\int_{0}^{v_{LMC}} v^2 f(\textbf{x},v) \, dv ,
\label{Eq:Chandrasekhar}
\end{multline}
where $m_{s}$ is the average mass of MW stars, $M_{LMC}$ is the mass of the LMC, $\textbf{v}_{LMC}$ is the velocity of the LMC with respect to the Galactic Center, and $f(\textbf{x},v)$ is the phase space distribution function of MW stars. We calculate the Coulomb Logarithm, $\ln(\Lambda)$, over the course of the simulation as in \cite{van2012,Patel2020}, that is:
\begin{equation}
\ln(\Lambda) = \max[L, \ln(r/C a_s)^{\alpha}],
\end{equation}
where $r$ is the LMC distance from the MW center, $L$ = 0, $C$ = 1.22, $\alpha$ = 1, and $a_s$ is the scale radius of the LMC.
 
The distribution function, $f(\textbf{x},v)$, is calculated for an arbitrary MW density in order to remain compatible with the various potential models in MilkyWay@home. We approximate the distribution at $\textbf{x}$ as a uniform density field with a Maxwellian velocity distribution, which is given by
\begin{equation}
f(\textbf{x}, v) = \frac{n(\textbf{x})}{(2 \pi \sigma^{2}(\textbf{x}))^{3/2}}e^{-\frac{v^{2}}{2 \sigma^{2}(\textbf{x})}},
\end{equation}
where $n(\textbf{x})$ is the number density at $\textbf{x}$, and $\sigma(\textbf{x})$ is the 1-dimensional velocity dispersion of MW stars at $\textbf{x}$. By plugging this distribution function into Eqn. \ref{Eq:Chandrasekhar}, we have
\begin{multline}
\textbf{F}_{DF} = -\frac{4\pi G^{2}\ln(\Lambda)\rho_{MW}(\textbf{x})M_{LMC}}{v_{LMC}^{3}} \\
\left[\operatorname{erf}  \left(\frac{v_{LMC}}{\sigma(\textbf{x})\sqrt{2}}\right)- \sqrt{\frac{2}{\pi}} \frac{v_{LMC}}{\sigma(\textbf{x})}e^{-\frac{v^{2}_{LMC}}{2 \sigma^{2}(\textbf{x})}}\right] \textbf{v}_{LMC},
\end{multline}
where $\rho_{MW}$ is the MW mass density.

Now, we need to find the value of the velocity dispersion, $\sigma(\textbf{x})$ in the DF. We approximate $\sigma(\textbf{x})$ by looking at the Spherical Jeans Equation, which is given by
\begin{multline}
\frac{\partial(\rho_{MW}(r)\sigma^{2}_{r})}{\partial r} + \frac{2 \rho_{MW}(r)}{r}(\sigma_{r}^{2} - \sigma_{\theta}^{2}) \\
+ \rho_{MW}(r) \frac{d\Phi_{MW}}{dr} = 0,
\label{Eq:Jeans}
\end{multline}
where $\Phi_{MW}$ is the gravitational potential of the MW. By assuming isotropy, $\sigma_{r}^{2} = \sigma_{\theta}^{2}$, we can simplify Eqn. \ref{Eq:Jeans} to

\begin{equation}
  \frac{\partial(\rho_{MW}(r)\sigma^{2}_{r})}{\partial r} = -\rho_{MW}(r) \frac{d\Phi_{MW}}{dr}. 
\label{Eq:Jeans_isotropy}
\end{equation}

Since $\textbf{a}_{MW} = -\nabla\Phi_{MW}$, Eqn. \ref{Eq:Jeans_isotropy} can be written as
\begin{equation}
  \frac{\partial(\rho_{MW}(r)\sigma^{2}_{r})}{\partial r} = \rho_{MW}(r) (\textbf{a}_{MW} \cdot \hat{r}). 
\label{Eq:Jeans_final}
\end{equation}
By integrating Eqn. \ref{Eq:Jeans_final} from the radius where our LMC resides ($r_{LMC}$) to infinity, we have
\begin{multline}
\rho_{MW}(r)\sigma^{2}_{r}|_{r=\infty} - \rho_{MW}(r)\sigma^{2}_{r}|_{r=r_{LMC}} \\
= \int_{r_{LMC}}^{\infty} \rho_{MW}(r) (\textbf{a}_{MW} \cdot \hat{r}) \, dr.
\end{multline}

If we look at larger radii, then the MW density drops to zero. Furthermore, for a gravitationally bound body to exist at larger radii, the velocity of the body must approach zero. This implies that its velocity dispersion also approaches zero. Therefore, we can approximate the velocity dispersion of MW stars at $r_{LMC}$ as
\begin{equation}
\sigma^{2}_{r_{LMC}} = -\frac{\int_{r_{LMC}}^{\infty} \rho_{MW}(r) (\textbf{a}_{MW} \cdot \hat{r}) \, dr}{\rho_{MW}(r_{LMC})}.
\end{equation}
We calculate this improper integral by setting the upper limit of integration to be 50 times the scale length of the MW halo. The DF is calculated given the position and velocity of the LMC. We can then find $\textbf{a}_{0,LMC}$ by adding the DF to the acceleration due to the MW. Note that the direction of the DF needs to be flipped in the backwards integration because this is a force that acts in opposition to the LMC's motion in the forward integration.

Because we cannot shift the MW position from the origin, we apply an opposite shift to the DG and LMC as 
\begin{equation}\label{eq:accel_shift}
\textbf{a}'_{t} = \textbf{a}_{t} - \textbf{a}_{t,MW}.
\end{equation}
The next positions of the DG and LMC can be found by these shifted accelerations as
\begin{equation}
    \textbf{v}_{t+\frac{1}{2}} = \textbf{v}_{t} + \textbf{a}'_{t}\frac{dt}{2},
\end{equation}
\begin{equation}
    \textbf{x}_{t+1} = \textbf{x}_{t} + \textbf{v}_{t+\frac{1}{2}}dt,
\end{equation}
where $dt$ is the time step in the simulation. Based on the updated positions, we find the updated accelerations experienced by the MW ($\textbf{a}_{t,MW}$), DG ($\textbf{a}_{t,DG}$), and LMC ($\textbf{a}_{t,LMC}$) in the same manner as before. Then, we again apply an opposite shift to these updated accelerations as Eqn. \ref{eq:accel_shift}. We find the velocities of the DG and LMC at this time step as
\begin{equation}
    \textbf{v}_{t+1} = \textbf{v}_{t+\frac{1}{2}} + \textbf{a}'_{t}\frac{dt}{2}.
\end{equation}
We store $\textbf{a}_{MW}$ for each time step in an array so we can later use it during forward integration. 

We use the final position of a point mass that represents the DG and the final position of a rigid LMC at the end of the backwards integration as the starting positions of the forward integration. At this position, we replace a point mass that represents the DG by its N-body representation in the same manner described in Section \ref{subsec:nbody}. 

We then apply essentially the same calculations as the backwards integration to the forward integration. The acceleration of each body in the DG is calculated as the sum of the acceleration due to the MW, the LMC, and each of the other bodies in the DG. The acceleration of the LMC is calculated as the sum of the acceleration due to the MW and DF. The $\textbf{a}_{MW}$ values are read from the end of the backwards array towards the present day, and the recorded $\textbf{a}_{MW}$ at that time step is subtracted from the acceleration of each body and the LMC so that the simulation stays in a frame of reference in which the center of the MW remains stationary at the center of the frame.

\begin{figure*}
    \centering
    \textbf{Best-fit OCS with Plummer LMC}\par\medskip
    \includegraphics[width=0.9\textwidth]
    {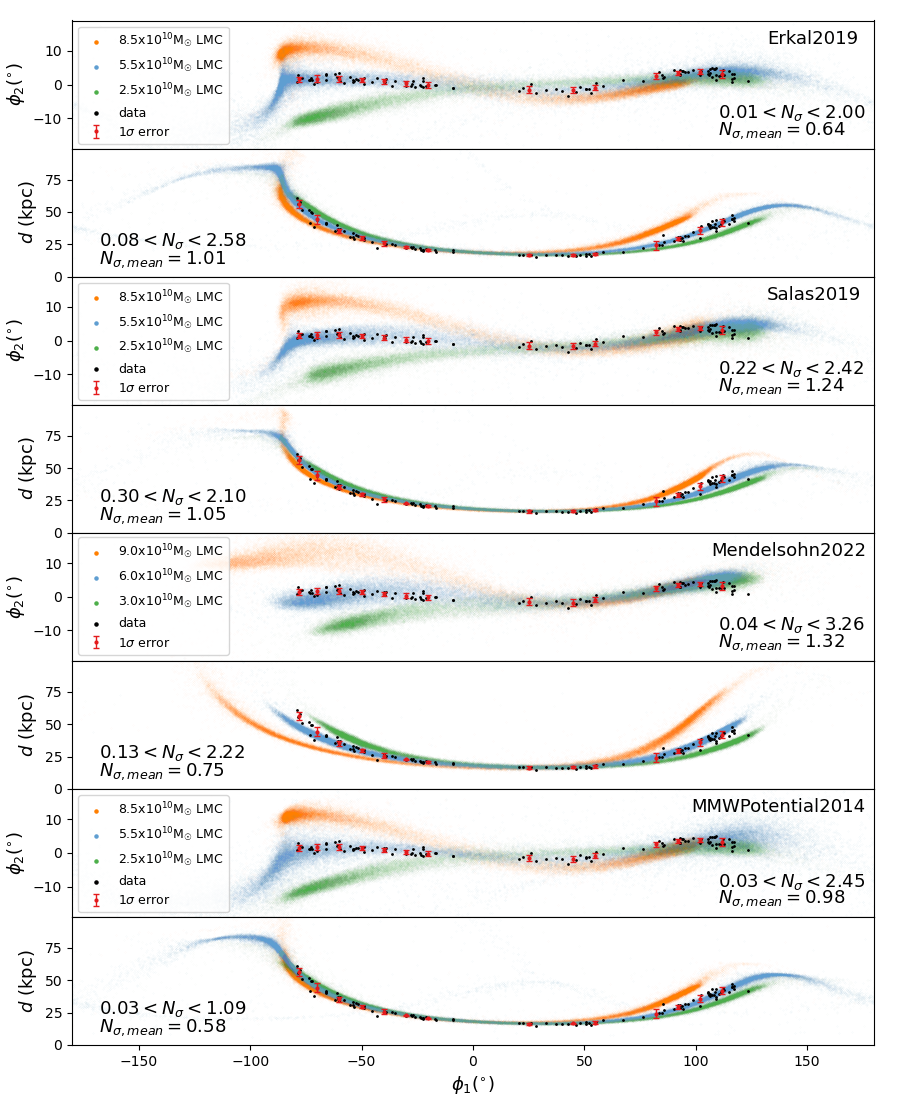}
    \caption{N-body simulations of the OCS with Plummer LMC with \textit{Erkal2019}, \textit{Salas2019}, \textit{Mendelsohn2022}, and \textit{MMWPotential2014} MW potential models (from top to bottom). In addition to our best-fit OCS (blue), we included streams with an LMC mass that differs from the best-fit mass by $\pm$ 3.0$\times 10^{10}$ M$_\odot$. The one with the higher LMC mass is colored with orange, and the lower mass is colored with green. The best-fit LMC mass using \textit{Erkal2019}, \textit{Salas2019}, and \textit{MMWPotential2014} potentials is M$_{LMC} = $ 5.5$\times 10^{10}$ M$_\odot$. The best-fit LMC mass using \textit{Mendelsohn2022} is M$_{LMC} = $ 6.0$\times 10^{10}$ M$_\odot$. Note that we only changed the LMC mass in 5.0$\times 10^{9}$ M$_\odot$ increments. The range and mean of deviations of the data points from the center of the simulation is given in the bottom right corner for the $\phi_2$ vs. $\phi_1$ panels and the bottom left corner for the $d$ vs. $\phi_1$ panels. Note that the simulated stream with the \textit{Mendelsohn2022} potential is not well fit to data; the stream center in $\phi_2$ vs. $\phi_1$ largely deviates from data near $\phi_1=-80^\circ$. In the $d$ vs. $\phi_1$ panel, it also misses data points at $\phi_1<-70^\circ$ although the 1$\sigma$ errors are higher for these data points. This slight discrepancy could mean that the \textit{Mendelsohn2022} is not as good a representation of the MW as the other three potentials, a suggestion that is supported by the rotation curves in Figure \ref{Fig:MW_potential}.}
    \label{Fig:OCS_fit_Plummer}
\end{figure*}

\begin{figure*}
    \centering
    \textbf{Proper motions of best-fit OCS with Plummer LMC}\par\medskip
    \includegraphics[width=0.9\textwidth]
    {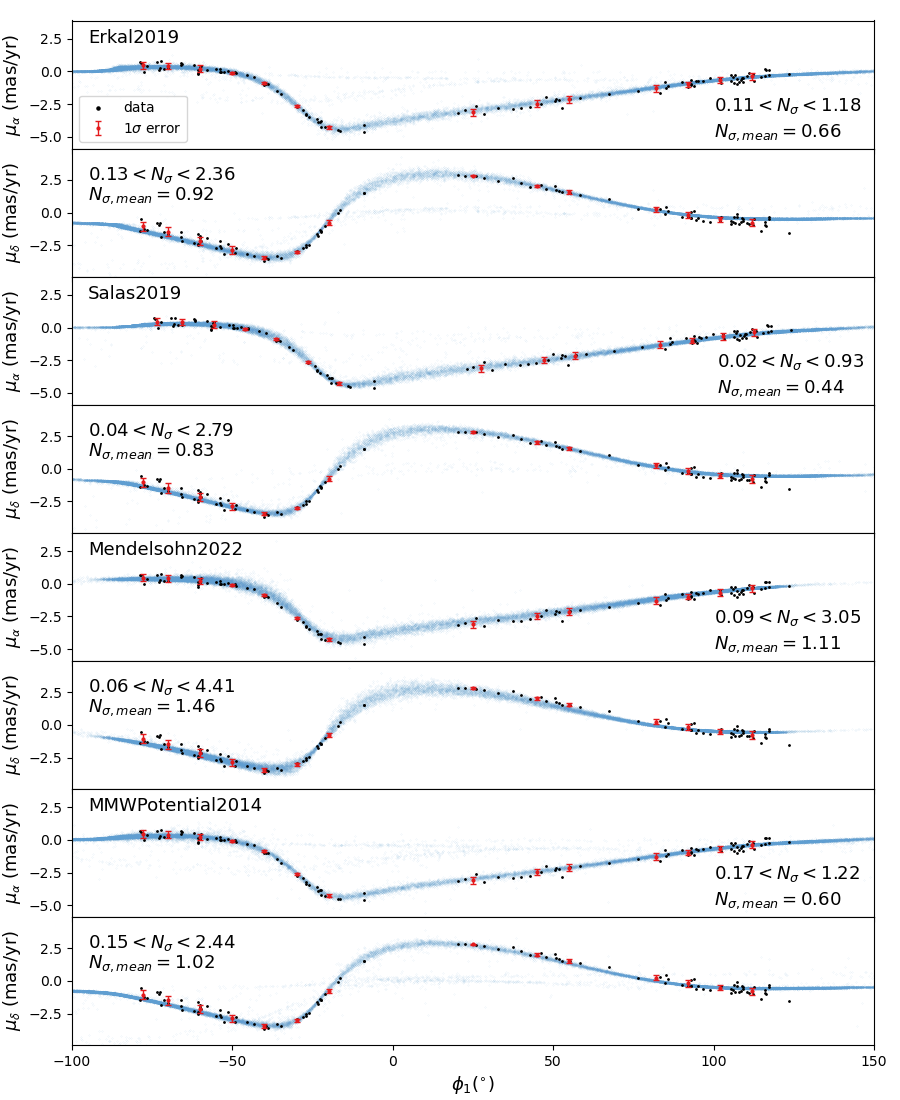}
    \caption{Proper motions in RA and Dec of our best-fit OC streams with Plummer LMC using \textit{Erkal2019}, \textit{Salas2019}, \textit{Mendelsohn2022}, and \textit{MMWPotential2014} MW potential models (from top to bottom). The black points are data from \cite{Koposov2019}. The range  and mean of deviations of the data points from the center of the simulation is given in the bottom right corner of the $\mu_{\alpha}$  vs. $\phi_1$ panels and the top left corner for the $\mu_{\delta}$  vs. $\phi_1$ panels. They all fit well, except that the proper motions for \textit{Mendelsohn2022} show a larger-than-expected error of 3-4$\sigma$.}
    \label{Fig:Plummer_PM}
\end{figure*}

\subsection{Fitting the Stream} \label{sec:fitting_stream}

We ran N-body simulations with MilkyWay@home as described in Section \ref{subsec:nbody}. We modeled the DG by a single Plummer sphere profile with 40,000 bodies, a scale radius of 1.0 kpc, and a progenitor mass of $10^{7}$ M$_\odot$. We chose to evolve the N-body simulation 3.6333 Gyr backward in the reverse orbit integration and then forward in time to match to the present stream position because this value was shown by \cite{Mendelsohn2022} to best fit the OCS. However, the evolution time of the simulation has little effect on the mass measurement. Increasing the evolution time will only make the simulated stream a little longer while making the stream a little narrower because it becomes more tidally disrupted over time.


Starting from the best-fit orbit parameters in \citetalias{Erkal2019} (spherical halo with the reflex motion) for \textit{Erkal2019} and \textit{MMWPotential2014}, and the best-fit orbit parameters from \cite{Mendelsohn2022} for \textit{Mendelsohn2022} and \textit{Salas2019}, we adjusted these parameters of the OCS as well as the mass of the LMC by hand in order to fit the $\phi_{1}$ and $\phi_{2}$ coordinates, the heliocentric distance, and the proper motions in RA and Dec of stars listed by  \cite{Koposov2019} as stream members. These stream members trace the path of the stream through the MW. Because the LMC is in the southern Galactic hemisphere, it significantly affects the southern part of the stream (left side of a stream in Figure \ref{Fig:nbody_particle_spray}) more than the northern part (right side of a stream in Figure \ref{Fig:nbody_particle_spray}), even though the LMC does affect both arms of the stream. As the LMC comes close to the OCS, its gravity bends the stream with a perturbation that is proportional to the LMC mass. By fitting the path, we can constrain the mass of the LMC. 

The mass of the LMC was first adjusted in 1.0$\times 10^{10}$ M$_\odot$ increments until the best match between the stream stars and the simulation bodies was achieved. The best-fit was assessed from $N_\sigma$ of $\phi_{2}$, $d$, $\mu_{\alpha}$, and $\mu_{\delta}$ (see Section \ref{sec:sigma_error}). Next, the orbit of the stream was adjusted by increasing or decreasing the present total velocity and thus the total energy of the stream. We shifted the starting position ($l$ and $b$) if the whole stream needed to tilt. The stream was fairly well fit to data by adjusting these four parameters one at a time until the best-fit was achieved. The velocity in each direction in the Galactic Cartesian coordinates $(V_{x}, V_{y} ,V_{z})$ was then adjusted in 1.0 kpc/Gyr increments, and then the mass of the LMC was adjusted in 5.0$\times 10^{9}$ M$_\odot$ increments, to adjust tilts at the tips of a stream. 

Fitting by hand in this way may not guarantee the best-fit using all parameters, meaning that it is possible that another set of parameters with different LMC mass can fit the data as well or better than our best-fit parameters. However, we did not have to change the orbit parameters much from \citetalias{Erkal2019} (which fitted the OCS parameters to the same data using a best likelihood algorithm) values when the same MW potential model was used as \citetalias{Erkal2019}. We will discuss how well our simulated OCS fits to data in the following section. 

\subsection{Results}

Figure \ref{Fig:OCS_fit_Plummer} shows in blue the stream fitted with two panels for each Milky Way potential model (from top to bottom: \textit{Erkal2019}, \textit{Salas2019}, \textit{Mendelsohn2022}, \textit{MMWPotential2014}). 
We additionally show how the mass of the LMC affects the path of the OCS by including streams with an LMC mass that differs from the
best-fit by $\pm$ 3.0$\times 10^{10}$ M$_\odot$. These additional streams are generated with the same orbit parameters, but with lower (green) and higher (orange) LMC mass during simulation. Our best-fit LMC
mass is M$_{LMC} = $ 5.5$\times 10^{10}$ M$_\odot$ for \textit{Erkal2019}, \textit{Salas2019}, and \textit{MMWPotential2014} and 6.0$\times 10^{10}$ M$_\odot$ for \textit{Mendelsohn2022}; these masses only differ by the increment used to search for the best-fit mass. Recall that the scale radius was selected for each LMC mass so that the mass within 8.7 kpc was 1.7$\times 10^{10}$ M$_\odot$. The best-fit parameters are summarized in Table \ref{tab:best_fit_params}. 

Surprisingly, the best-fit mass estimates are less than half of the estimations found by many other researchers (see Section \ref{sec:intro}), including those that are made by fitting exactly this tidal stream. In this analysis, we modeled the LMC as a Plummer sphere, which is not a popular model for the LMC. Since the NFW profile describes dark matter halos of galaxy clusters \citep{Navarro1997}, the LMC is often modeled as the NFW \citep{Vasiliev2021,Magnus2022,Koposov2023} or Hernquist profile \citep{Besla2012,Laporte2018,Erkal2019,Erkal2020,Shipp2021}, which is similar to the NFW in the inner region. In Section \ref{sec:LMC_model}, we compare the fit with a Plummer LMC to the fit with a Hernquist LMC.

The best-fit LMC mass did not depend strongly on which MW potential model was used. This result is likely a result of the MW potential being well constrained in the range of Galactic distances the OCS travels through when the LMC gets close to it. 

Figure \ref{Fig:Plummer_PM} shows the proper motions in RA and Dec of our best-fit streams under all four different MW potential models. The number of sigma, $N_\sigma$, for each parameter is presented in each panel of Figure \ref{Fig:OCS_fit_Plummer} and \ref{Fig:Plummer_PM}. They all fit well, except for the \textit{Mendelsohn2022} case. We found fitting the stream with \textit{Mendelsohn2022} to be more challenging, and this fit was not as good as other MW potential cases. It is possible that this potential was difficult to fit because it does not represent the actual MW potential; recall that it produced a rotation curve (Figure \ref{Fig:MW_potential}) that did not match the measured MW rotation curve as well as the other three. 


\begin{figure*}
    \includegraphics[width=\textwidth]{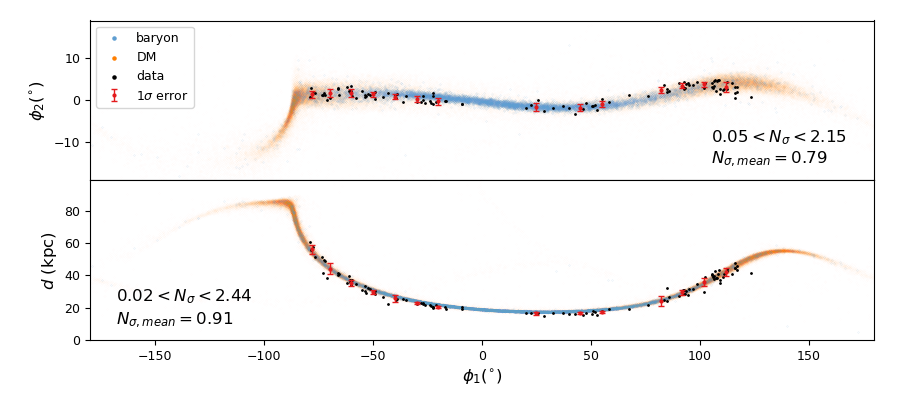}
    \caption{Simulated OCS derived by modeling the DG with a two-component Plummer sphere that represents baryon particles and dark matter particles separately. We switched the DG model from a single Plummer to a double Plummer here, but we used the same best-fit OCS parameters using the \textit{Erkal2019} potential with a Plummer model LMC listed in Table \ref{tab:best_fit_params}. The baryon particles are colored with blue, and the dark matter particles are colored with orange. This illustrates the long tidal tails that exceed the data in Figure \ref{Fig:OCS_fit_Plummer} are mostly dark matter particles. The fact that the baryon (blue) particles are still a good match to the real data suggests that use of a one or two component DG does not significantly affect the LMC mass measurement.}
    \label{Fig:double_plummer}
\end{figure*}



We note that the number of sigma in each parameter is statistical only, and does not include systematic errors. The statistical errors in $d$, $\mu_{\alpha}$, and $\mu_{\delta}$ are much smaller than the systematic errors in the observed data. We expect systematic error in the observed distance to be about 3$\%$ of that distance \citep{Li2023}. Since the OCS is roughly 20 kpc away from us, the systematic error is about 0.6 kpc. If we add this systematic error to the statistical error for the \textit{Erkal2019} case, then 0.08 $< N_{\sigma_{d}} <$ 1.81 for all of the measured stream positions. The systematic errors in observed proper motions are about 0.1 mas/yr \citep{Luri2018}. If these systematic errors are added to the statistical error, then, we calculate 0.10 $< N_{\sigma_{\mu_{\alpha}}} <$ 1.11 and 0.11 $< N_{\sigma_{\mu_{\delta}}} <$ 2.03. All of these deviations indicate the fit is very good.

On the other hand, the star position uncertainties of Gaia are very small \citep{Gaia_Collaboration2018}. In $\phi_{2}$, the largest errors are from the range 80$^{\circ}$ $<\phi_1 <$ 100$^{\circ}$, where the simulation is off by about 1$^{\circ}$ from the data. This stream data set has its own systematic errors as these stars were selected in a 4$^{\circ}$-wide stream track on the sky \citep{Koposov2019}. In particular, $\phi_1 =$ 90$^{\circ}$ is where the stars begin to be contaminated with the background stars or other structures. If a few background stars leak into the stream sample, the apparent position of the stream will be affected. Since stars that are farthest from the stream track are cut out of the sample, the standard deviation of the selected stars will be systematically smaller. In addition, we do not include any information about the substructures and deformation of the halo in our Galactic potential. For example, \cite{Bonaca2019} showed that a small perturbation, such as from dark matter subhalos, can cause a spur of stars separated by about 2$^{\circ}$ from stream. The deformation of the MW is also known to influence the OCS path \citep{Lilleengen2023}. Any missing information in the Galactic potential can lead to a shift in the stream path. Moreover, since there are systematic errors in other parameters, for example if the estimated distance to the stream stars is too small by 3$\%$, then it would change calculation of the stream orbit and could change the shape of the orbit on the sky. In fact, a simulated OCS by \citetalias{Erkal2019} (Figure \ref{Fig:nbody_particle_spray}) shows similar deviations from data around 80$^{\circ}$ $<\phi_1 <$ 100$^{\circ}$, even though this paper found the best-fit by using a best likelihood algorithm. Therefore, we believe the fitting algorithm, even though it was done by hand, is sufficient.

\section{Effect of Using a Two-Component Dwarf Galaxy Model} \label{sec:two_component} 

Here, we switch the model of the DG from a single Plummer to a double Plummer that includes both stellar and dark matter components, and show that the OCS fit does not change. We selected the DG progenitor parameters from \cite{Mendelsohn2022}: a baryon scale radius of 0.1812 kpc, a baryon mass of 2.7$\times$ 10$^{5}$ M$_\odot$, a dark matter scale radius of 0.8100 kpc, and a dark matter mass of 2.1$\times$ 10$^{7}$ M$_\odot$. The initial DG was created using the methods outlined in \cite{Shelton2018, Shelton2021, Mendelsohn2022}. We kept the other parameters the same as the one component model; all other parameters are as listed in the P-Erkal2019 line of Table \ref{tab:best_fit_params}.

Figure \ref{Fig:double_plummer} shows the best-fit OCS using the \textit{Erkal2019} potential, a Plummer model LMC, and a model DG that is composed of 10,000 baryon particles and 78,257 dark matter particles, both distributed in space according to Plummer profiles. The number of star particles was selected so there were enough to sample the stream well. The number of dark matter particles was calculated based on the fact that there is 78 times as much dark matter and the mass per particle must not more than 10 times the star mass per particle to avoid DF effects. The baryon particles are in blue, and the dark matter particles are in orange in the figure. By comparing Figure \ref{Fig:double_plummer} with the top two panels in Figure \ref{Fig:OCS_fit_Plummer}, we see that the long tidal tails that extend past the data in Figure \ref{Fig:OCS_fit_Plummer} are mostly dark matter particles, and the path of the stream remains the same as for the simulation with a single Plummer DG profile. 

For the rest of this paper, we modeled the DG as a single Plummer instead of a two-component model because it does not affect the LMC mass measurement, and the thicker stream width produced by a single Plummer model is easier to compare with \citetalias{Erkal2019}, who also modeled the DG with a single Plummer sphere profile.

\begin{figure*}
    \centering
    \textbf{Best-fit OCS with Hernquist LMC}\par\medskip
    \includegraphics[width=0.9\textwidth]{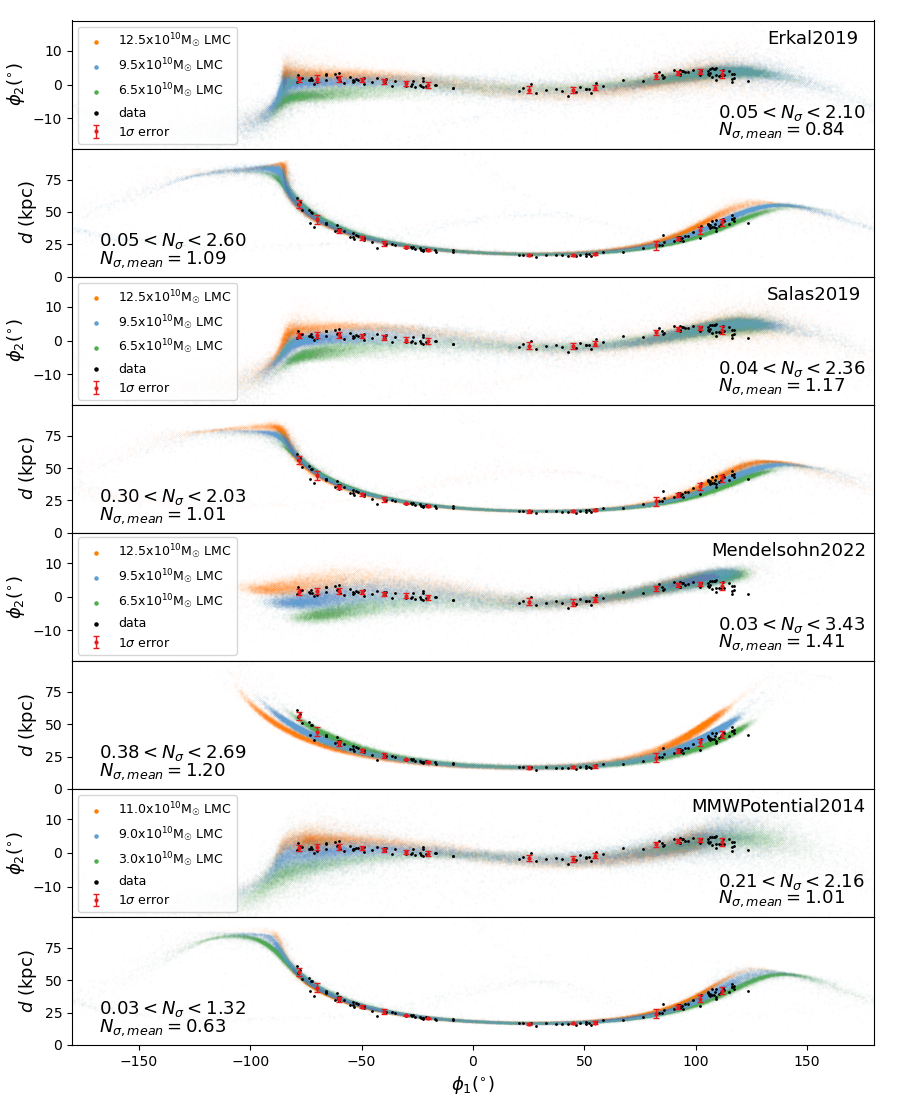}
    \caption{Stream-fit with Hernquist LMC under the gravitational influence of four different model MW potentials: \textit{Erkal2019}, \textit{Salas2019}, \textit{Mendelsohn2022}, and \textit{MMWPotential2014}  (from top to bottom). In addition to our best-fit OCS (blue), we included streams with the LMC of $\pm$ 3.0$\times 10^{10}$ M$_\odot$ (orange and green). We obtained M$_{LMC} = $ 9.5 $\times 10^{10}$ M$_\odot$ for \textit{Erkal2019}, \textit{Salas2019} and \textit{Mendelsohn2022}, and 9.0$\times 10^{10}$ M$_\odot$ for \textit{MMWPotential2014}. The range and mean of deviations of the data points from the center of the simulation is given in the bottom right corner for the $\phi_2$ vs. $\phi_1$ panels and the bottom left corner for the $d$ vs. $\phi_1$ panels. Note that the simulated stream with the \textit{Mendelsohn2022} potential is again not well fit to data. The stream center in $\phi_2$ vs. $\phi_1$ largely deviates from data near $\phi_1\backsim-80^\circ$. In $d$ vs. $\phi_1$, it also misses data points at $\phi_1<-80^\circ$ although the 1$\sigma$ errors are higher for these data points. This slight discrepancy could be because this potential differs from the other three in the Galactocentric radius range $5<r<20$ kpc, as shown in Figure \ref{Fig:MW_potential}.}
    \label{Fig:OCS_fit_Hernquist}
\end{figure*}

\begin{figure*}
    \centering
    \textbf{Proper motions of best-fit OCS with Hernquist LMC}\par\medskip
    \includegraphics[width=0.9\textwidth]
    {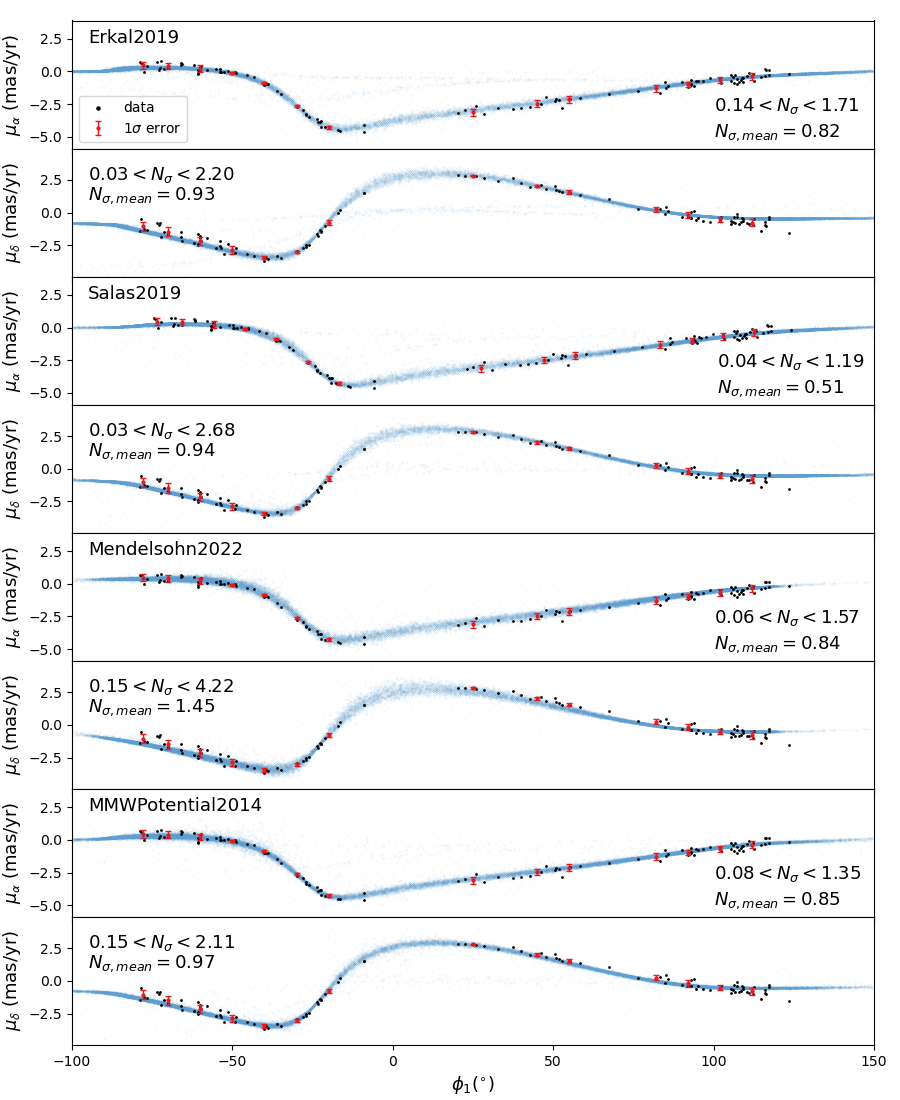}
    \caption{Proper motions in RA and Dec of our best-fit OC streams with Henrquist LMC under \textit{Erkal2019}, \textit{Salas2019}, \textit{Mendelsohn2022}, and \textit{MMWPotential2014} MW potential models (from top to bottom). The black points are data from \cite{Koposov2019}. The range and mean of deviations of the data points from the center of the simulation is given in the bottom right corner for the $\mu_{\alpha}$  vs. $\phi_1$ panels and top left corner for the $\mu_{\delta}$  vs. $\phi_1$ panels. Again, they all fit well, except that the proper motion in Dec for \textit{Mendelsohn2022} deviates from the data by as much as 4$\sigma$, which is significantly larger than any other model.}
    \label{Fig:hernquist_PM}
\end{figure*}
 
\begin{figure}
    \centering
    \includegraphics[width=0.45\textwidth]{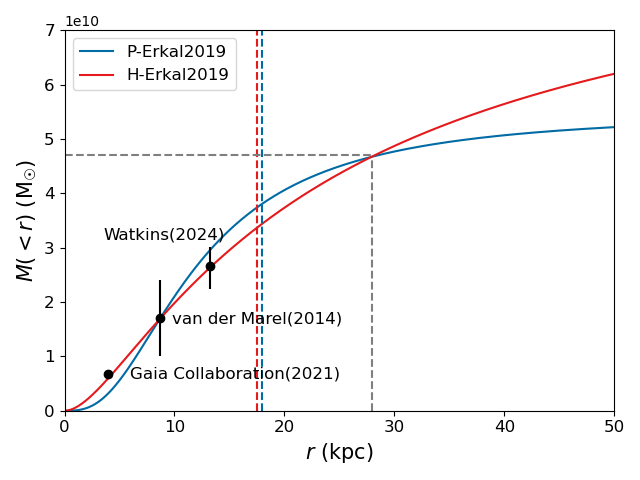}
    \caption{Enclosed mass of our best-fit Plummer LMC (blue) and Hernquist LMC (red). The dashed vertical lines show the closest distance from the simulated OCS to the LMC for each profile. The black dots are the enclosed masses from \cite{Van2014}, which we used to calculate the scale radius of the LMC, and \cite{Watkins2024} (M(13.2kpc)=2.66$^{+0.42}_{-0.36}$  $\times 10^{10}$ M$_\odot$). We also calculated the enclosed mass from \cite{Gaia_Collaboration2018} (M(4kpc)=6.72$\pm$ 0.40  $\times 10^{9}$ M$_\odot$). The gray dashed line at 28 kpc shows the point where both values agree again with M$(<r)$ = 4.7 $\times 10^{10}$ M$_\odot$. For the same mass within 28 kpc, the Hernquist profile has a much larger total mass (at large radius).}
    \label{Fig:enclosed_mass_best_fit}
\end{figure}

\begin{figure}
    \centering
    \includegraphics[width=0.45\textwidth]{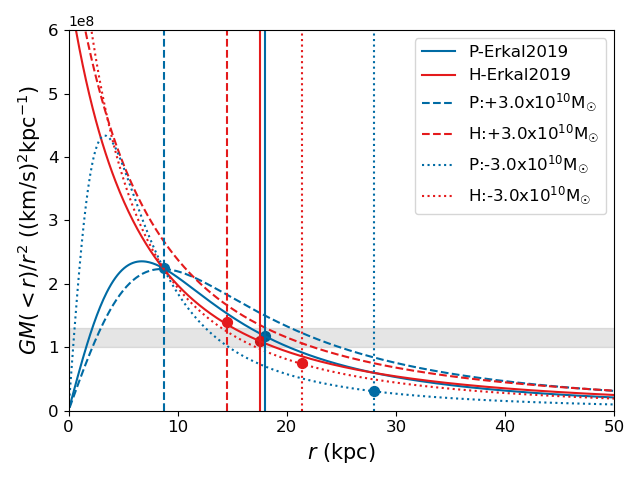}
    \caption{Acceleration of the LMC as a function of its radius for each OCS progenitor model in Figure \ref{Fig:OCS_fit_Plummer} and Figure \ref{Fig:OCS_fit_Hernquist} (solid lines shows the best-fit LMC, and dashed and dotted lines show the LMC of $\pm$ 3.0$\times 10^{10}$ M$_\odot$ with Plummer profile, blue, and Hernquist profile, red). The vertical lines show the closest distance from the simulated OCS to the LMC for each profile using the \textit{Erkal2019} MW potential, and the large dots show the acceleration at closest approach for each case. Note that the blue (Plummer) dots have a much larger difference in acceleration at closest approach than the red (Hernquist) dots do for the same change in LMC mass. Also, note that the best-fit Plummer and Hernquist profiles here have similar acceleration at closest approach.}
    \label{Fig:LMC_model_accel}
\end{figure}

\section{Effect of Particle-Spray method on LMC mass Measurement} \label{sec:simulation_method}

In order to directly compare with the \citetalias{Erkal2019} particle-spray simulation, we modeled the LMC as a Hernquist profile instead of a Plummer profile as was used in Section \ref{sec:potential}. Again, we ran N-body simulations with MilkyWay@home and used the same progenitor parameters under the gravitational influence of four different MW potential profiles. As with the Plummer profile, the scale radius of the Hernquist profile LMC was selected so that the mass within 8.7 kpc is 1.7$\times 10^{10}$ M$_\odot$. We fitted the simulated OCS to data in the same manner as in Section \ref{sec:fitting_stream}, starting from the best-fit orbit parameters in Table \ref{tab:best_fit_params} for the Plummer LMC cases.

Figure \ref{Fig:OCS_fit_Hernquist} shows the fitted OCS under each MW potential model (from top to bottom: \textit{Erkal2019}, \textit{Salas2019}, \textit{Mendelsohn2022}, \textit{MMWPotential2014}). As before, we included streams that were generated with an LMC that differs from the best-fit LMC mass by $\pm$ 3.0$\times 10^{10}$ M$_\odot$ (orange and green) using the same orbit parameters as our best-fit OCS (blue) for each MW potential model to show how the mass of the LMC affects the path of the OCS. 

We obtained M$_{LMC} = $ 9.5 $\times 10^{10}$ M$_\odot$ for \textit{Erkal2019}, \textit{Salas2019}, and \textit{Mendelsohn2022} and  9.0$\times 10^{10}$ M$_\odot$ for \textit{MMWPotential2014}. The best-fit parameters are summarized in Table \ref{tab:best_fit_params}. Figure \ref{Fig:hernquist_PM} shows the proper motions in RA and Dec of our best-fit streams under all four different MW potential models. Just as was shown in Figure \ref{Fig:Plummer_PM}, the simulations all fit the proper motion data well, except for the \textit{Mendelsohn2022} case.

We point out that we again find that the mass measurement is not much affected by the MW potential models, as was the case for Plummer profile LMC simulations in Section \ref{sec:potential}. The result of the mass estimate with \textit{Erkal2019} here implies that by using the N-body simulation instead of the particle-spray modeling, the estimated mass of the LMC decreases by 3.0 $\times 10^{10}$ M$_\odot$; this simulated stream was created using the same MW potential with the same LMC model (a Hernquist density profile) as \citetalias{Erkal2019}. This shows that significant systematic errors can be introduced by the use of the particle-spray technique.

We note that we did not try all available particle-spray models (see Section \ref{sec:simulation}) to compare with our N-body simulation results. The particle-spray modeling technique has been improving over time \citep{Chen2024}, and a newer version of particle-spray could reduce or eliminate the error introduced by particle-spray approximation.

\section{Effect of the LMC Density Profile} \label{sec:LMC_model}

Here, we show that the assumed LMC radial profile has a large influence on the estimate of the total LMC mass. In particular, we discuss the following scenarios: modeling the LMC different density profiles, varying its scale radius, and cutting off the density profile at its tidal radius.

\subsection{Hernquist vs. Plummer} \label{sec:Hernquist_vs_Plummer}

The results in Section \ref{sec:simulation_method} show that just by switching the LMC model from a Plummer to a Hernquist profile, the total mass of the LMC needed to be increased by 3.5 - 4.0 $\times 10^{10}$ M$_\odot$ to produce similar streams. We also note that changing the LMC mass by 3.0 $\times 10^{10}$ M$_\odot$ makes a bigger difference to the path of the OCS if the assumed LMC is a Plummer sphere (Figure \ref{Fig:OCS_fit_Plummer}) than it does if the LMC is a Hernquist profile (Figure \ref{Fig:OCS_fit_Hernquist}). The reason for this can be derived from Figure \ref{Fig:enclosed_mass_best_fit}.


\begin{figure*}
    \centering
    \textbf{Best-fit OCS with varying LMC scale radius}\par\medskip
    \includegraphics[width=0.9\textwidth]{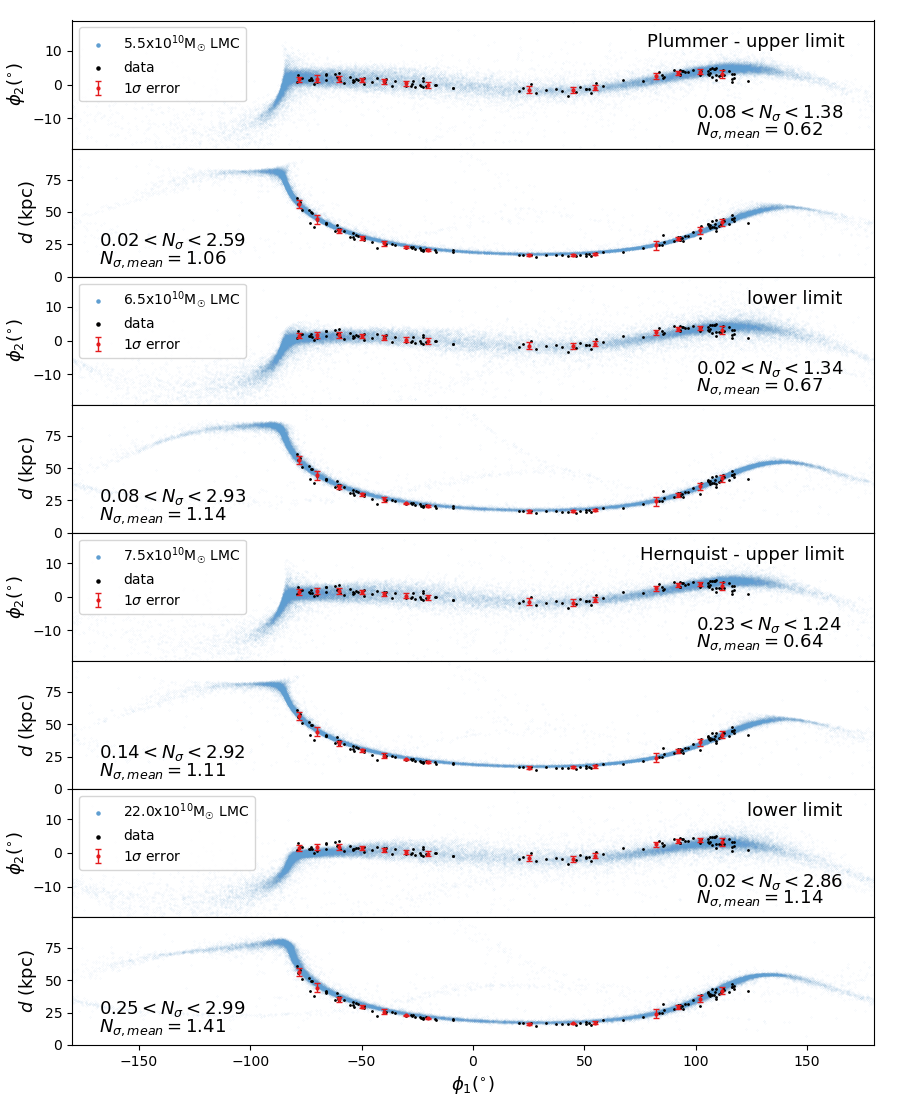}
    \caption{Best-fit OCS, when varying the mass enclosed within 8.7 kpc according to the upper and lower uncertainty bounds \citep[1.7 $\pm$ 0.7$\times 10^{10}$ M$_\odot$ according to][]{Van2014}. From top to bottom, the panels show the best-fit Plummer LMC using the upper bound mass within 8.7 kpc, the best-fit Plummer LMC using the lower bound mass within 8.7 kpc, the best-fit Hernquist LMC using the upper bound mass within 8.7 kpc, and the best-fit Hernquist LMC using the lower bound mass within 8.7 kpc. We fitted the simulated OCS to data using the \textit{Erkal2019} MW potential, and the best-fit parameters are listed in Table \ref{tab:best_fit_params}. The Plummer streams can be compared with the top two panels of Figure \ref{Fig:OCS_fit_Plummer}, and the Hernquist streams can be compared with the top two panels of Figure \ref{Fig:OCS_fit_Hernquist}. The range and mean of deviations of the data points from the center of the simulation is given in the bottom right corner for the $\phi_2$ vs. $\phi_1$ panels and the bottom left corner for the $d$ vs. $\phi_1$ panels. With the Hernquist LMC, the estimated LMC mass is more sensitive to the value of M($<$8.7kpc); note the change in central mass changes the LMC mass from 7.5$\times 10^{10}$ M$_\odot$ to 2.2$\times 10^{11}$ M$_\odot$ for the Hernquist LMC, and from 5.5$\times 10^{10}$ M$_\odot$ to 6.5$\times 10^{10}$ M$_\odot$ for the Plummer LMC. In both cases, the scale radius of the profiles were adjusted to fit M($<$8.7kpc), so the different mass estimates can equivalently be thought of as a consequence of changing the scale radius, or of changing M($<$8.7kpc).}
    \label{Fig:scale_radius}
\end{figure*}

\begin{figure}
    \centering
    \includegraphics[width=0.45\textwidth]{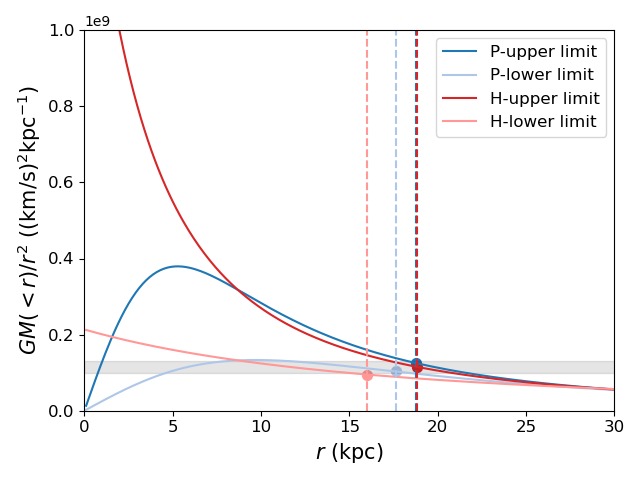}
    \caption{Acceleration of the best-fit LMC using the upper bound (dark color) and lower bound (light color) with the Plummer (blue) and Henrquist profile (red). The vertical dashed lines are the closest approach of the LMC, and the large dots show the acceleration at closest approach for each case. As for all other best-fit LMC masses, the acceleration is about 1.0--1.3 $\times$ 10$^{8}$ (km/s)$^2$kpc$^{-1}$.}
    \label{Fig:LMC_mosel_accel_limit}
\end{figure}

\begin{figure*}
    \centering
    \includegraphics[width=1.0\textwidth]{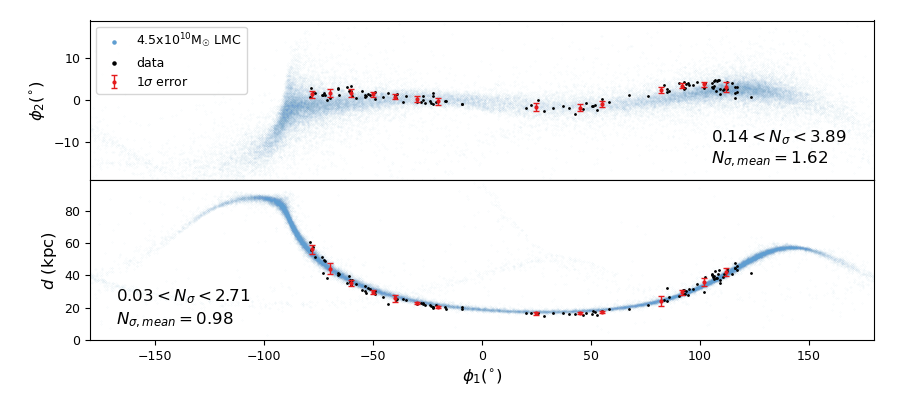}
    \caption{Fitted OCS with the Hernquist LMC with a cutoff. We selected the tidal radius of the LMC as the cutoff radius. The mass of the LMC in the simulation was set so that the mass inside of the tidal radius would be the total mass of the LMC while the mass within 8.7 kpc is 1.7 $\pm$ 0.7$\times 10^{10}$ M$_\odot$ \citep{Van2014}. We fitted the simulated OCS to data using the \textit{Erkal2019} MW potential, and its best-fit parameters are listed in Table \ref{tab:best_fit_params}. We obtained a mass of M$_{LMC}$($<$16.4 kpc) = 4.5$\times 10^{10}$ M$_\odot$. This is surprisingly consistent with the enclosed mass found from the two different LMC models (Figure \ref{Fig:enclosed_mass_best_fit}). The range and mean of deviations of the data points from the center of the simulation is given in the bottom right corner for the $\phi_2$ vs. $\phi_1$ panel and the bottom left corner for the $d$ vs. $\phi_1$ panel.}
    \label{Fig:LMC_cutoff}
\end{figure*}

Figure \ref{Fig:enclosed_mass_best_fit} shows the enclosed mass of our best-fit Plummer LMC (blue) and Hernquist LMC (red). The dashed vertical lines (at 18.0 and 17.5 kpc) are the closest approach of the LMC in both simulations. Here, the closest approach was determined by first calculating the center of the simulated stream at each time step, and then finding the distance from the center of mass of the LMC to the nearest stream center along the stream. The black dots are the enclosed masses within 8.7 kpc from \cite{Van2014}, which we used to calculate the scale radius of the LMC. The best-fit Plummer and Hernquist profiles have the same mass within this radius by construction. We also included the mass within 13.2 kpc (2.66$^{+0.42}_{-0.36}$  $\times 10^{10}$ M$_\odot$) from \cite{Watkins2024}, and we calculated the enclosed mass within 4 kpc (6.72$\pm$ 0.40  $\times 10^{9}$ M$_\odot$) from \cite{Gaia_Collaboration2018}. 

What is more interesting is that these two models agree once again at 28 kpc with M$_{LMC} (<r) = $ 4.7 $\times 10^{10}$ M$_\odot$. We will return to this later. 

For now, notice that the Hernquist mass profile continues to rise significantly even out at 50 kpc. A large part of the LMC mass is way outside of the distance of closest approach, where the LMC has the strongest effect on the stream. The mass that is way outside the interaction distance will have no effect on the stream because both the core of the LMC and the stream are inside this spherical distribution. With a Hernquist profile, we need more total mass to produce the same stream defection because a larger fraction of the mass is at large radii. Changing the LMC mass by 3.0 $\times 10^{10}$ M$_\odot$ makes a smaller difference for the Hernquist profile because a larger part of that extra mass does not participate in the deflection.

Another way to visualize the same information is shown in Figure \ref{Fig:LMC_model_accel}, which shows the acceleration of gravity due to the LMC as a function of radius from its center for the two best-fit LMC profiles. The solid lines are the best-fit profiles, and the dashed and dotted lines are for an LMC mass that differs from the best-fit by $\pm$ 3.0$\times 10^{10}$ M$_\odot$. The acceleration at closest approach changes more for the Plummer than the Hernquist profile for the same change in LMC mass. This explains the more significant differences in the simulated stream for the Plummer than the Hernquist profile when the mass in changed.  

\subsection{Scale Radius of the LMC Model} \label{sec:scale_radius}

In this section, we discuss how much the uncertainty of the LMC scale radius affects the LMC mass estimate as determined by the path of the OCS. In Section \ref{sec:potential} and Section \ref{sec:Hernquist_vs_Plummer}, the scale radius was selected so that the enclosed mass of the LMC within 8.7 kpc is 1.7 $\pm$ 0.7$\times 10^{10}$ M$_\odot$ \citep{Van2014}. Here, we vary the mass enclosed within 8.7 kpc  according to the upper and lower uncertainty bounds, and adjust the scale radius  accordingly. 

In this test, we used the \textit{Erkal2019} potential and the same progenitor parameters (but with an adjusted scale radius), evolved with MilkyWay@home N-body simulations. Using the scale radii calculated from the updated enclosed LMC masses, we recreated the streams with both Plummer and Hernquist LMC models. We again fitted the simulated OCS to data in the same manner as in Section \ref{sec:fitting_stream}.

Figure \ref{Fig:scale_radius} shows the fitted OCS with upper bound Plummer LMC, lower bound Plummer LMC, upper bound Hernquist LMC, and lower bound Hernquist LMC (from top to bottom). We obtained 5.5 $\times 10^{10}$ M$_\odot$ for the upper bound Plummer LMC, 6.5 $\times 10^{10}$ M$_\odot$ for the lower bound Plummer LMC, 7.5 $\times 10^{10}$ M$_\odot$ for the upper bound Hernquist LMC, and 2.2 $\times 10^{11}$ M$_\odot$ for lower bound Hernquist LMC. The best-fit parameters are summarized in Table \ref{tab:best_fit_params}. Note that the best-fit mass varies by a factor of four. Although the upper and lower uncertainty bounds have only a small effect on the estimated mass of the Plummer LMC, they have a large effect on the mass of the Hernquist LMC. The reason is the same as we found in Section \ref{sec:Hernquist_vs_Plummer}; the Hernquist profile has a larger fraction of the mass at large radii.

In Figure \ref{Fig:LMC_mosel_accel_limit}, we show the acceleration of the LMC using the upper bound (dark color) and lower bound (light color) with the Plummer (blue) and Henrquist (red) profile. The vertical lines show the closest approach of the LMC.

\begin{figure}
    \centering
    \includegraphics[width=0.45\textwidth]{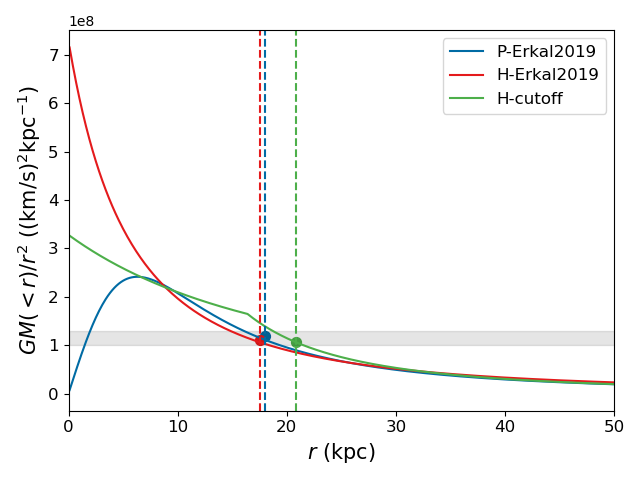}
    \caption{Gravitational acceleration as a function of radius for the best-fit cutoff Hernquist profile (green), as well as the full Plummer (blue) and Hernquist (red). The vertical dashed lines are the closest approach of the LMC to the OCS, and the large dots show the acceleration at closest approach for each case. Note that all these models have about the same acceleration at closest approach.}
    \label{Fig:LMC_model_accel_cutoff}
\end{figure}

\begin{figure*}
    \centering
    \includegraphics[width=0.6\textwidth]{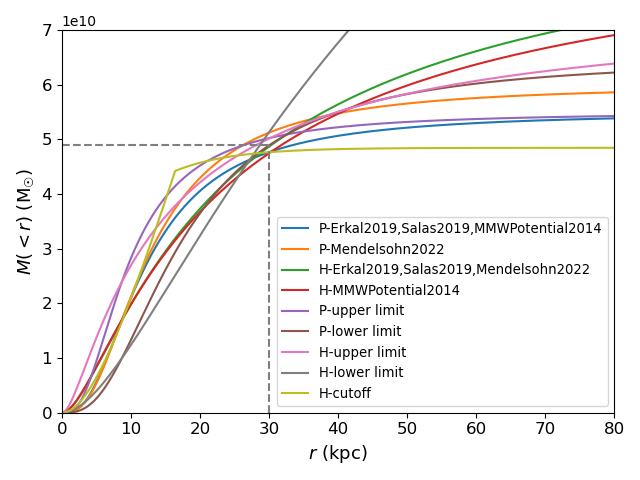}
    \caption{LMC enclosed mass vs. radius for all of the best-fit LMC profiles. The best-fit Plummer and Hernquist models in Table \ref{tab:best_fit_params} are labeled along with their MW potential names. Plummer/Hernquist-upper/lower bound models are explained in Section \ref{sec:scale_radius}, and Hernquist-cutoff model is described in Section \ref{sec:LMC_model_cutoff}. This shows all of the models cross each other near 30 kpc, where the mass enclosed is M($<r$) = $4.93 \times 10^{10}$ M$_\odot$.}
    \label{Fig:all_mass}
\end{figure*}

Note that, as before, the best-fit LMC profiles have about the same acceleration at the closest approach to the OCS. In order to put a smaller mass within 8.7 kpc for the lower bounds, while still keeping the same mass within $ \backsim$30 kpc, the scale radius had to increase dramatically, so that it is much larger than the distance of closest approach. With larger scale radius than this, a very large fraction of the increase in total mass is at very large radius, where the OCS is not affected. In fact, the lower bound Hernquist LMC model did not show much difference in the stream path after increasing its mass to above 2.0 $\times 10^{11}$ M$_\odot$, indicating there is no upper bound on the total mass of a Hernquist profile LMC. For the large ($>$ 30 kpc scale radius) model, the acceleration is nearly flat as a function of distance from the LMC. Larger mass Hernquist LMCs need to fit the fixed acceleration at 8.7 kpc, and therefore the scale radius must increase and all of the extra mass comes in at large radius where it has negligible effect on the OCS. We conclude that the radial shape of the LMC is not well constrained by the OCS.
 
\subsection{Cutoff Radius of the LMC Halo} \label{sec:LMC_model_cutoff}

To further explore the effect of the radial profile on the LMC mass estimate, we tried fitting the mass of the LMC under the assumption that its profile is a spherical Hernquist profile that is truncated at the tidal radius. It is reasonable to assume the dark matter in the LMC is in the process of being tidally stripped as it falls into the MW. While the actual profile of a tidally disrupting DG is quite complex, this is a sensible idealized approximation that is easy to explore. A Hernquist profile with the cutoff at large radius is given by:

\begin{equation}
\rho_{H}(r) =
    \begin{cases}
        \frac{aM}{2\pi} \frac{1}{r(a+r)^{3}} & \text{$r  \leqq b$}\\
        \rho^{*}\left( \frac{b}{r}\right)^{2} e^{-\frac{2r}{b}} & \text{$r > b$}\\
    \end{cases}
\end{equation}
\citep{Hwang2013}, where $a$ is the scale radius of the LMC, $M$ is the mass of the LMC, $b$ is the radius at which the cutting-off starts, and $\rho^{*}$ is chosen to ensure the continuity of $\rho_{H}(r)$ at $r$ = $b$.

Again, we used the \textit{Erkal2019} MW potential, and we selected the tidal radius of the LMC as the cutoff radius. The tidal radius was calculated using Eqn. \ref{Eq:tidal_radius}, and it changes as we adjust the mass of the LMC. The mass of the LMC in the simulation was set so that the mass inside of the tidal radius would be the total mass of the LMC while the mass within 8.7 kpc is fixed at 1.7 $\times 10^{10}$ M$_\odot$. When calculating the DF of the LMC, we used the tidal radius for the Coulomb Logarithm calculation instead of its scale radius since the tidal radius was much smaller than the scale radius. As in previous sections, we fitted the simulated OCS to the data in the same manner as in Section \ref{sec:fitting_stream}.

In Figure \ref{Fig:LMC_cutoff}, we show the best-fit OCS using the Hernquist LMC model with a cutoff, which has a mass of M$_{LMC}$($<$16.4 kpc) = 4.5$\times 10^{10}$ M$_\odot$. The best-fit parameters are listed in Table \ref{tab:best_fit_params}. Since there is essentially no LMC mass further than 16.4 kpc due to the cutoff in this model, this is also the total LMC mass. It is surprisingly consistent with the enclosed mass of 4.7$\times 10^{10}$ M$_\odot$ within 28 kpc found from the two LMC models in Figure \ref{Fig:enclosed_mass_best_fit}. 

Figure \ref{Fig:LMC_model_accel_cutoff} shows the acceleration of the best-fit LMC (green). We also included best-fit Plummer (blue) and Hernquist without cutoff (red) models in this plot to show how the cutoff Hernquist model differs from the other two. Although the cutoff Hernquist model shows a much smaller acceleration than the Hernquist model without a cutoff near the center, all models show approximately the same acceleration at their closest approaches, which are also around the same position from the center. It appears that an acceleration of 1.0--1.3 $\times$ 10$^{8}$ (km/s)$^2$kpc$^{-1}$ at closest approach is what is needed to produce the observed bend in the OCS, independent of the assumed radial profile of the LMC.

Note that our cutoff radius is more reasonable than previous estimates. \cite{Koposov2023} found a best-fit cutoff LMC radius of over 150 kpc when optimizing parameters by the likelihood scores in order to fit their OCS. This is approximately 3 times the distance from the center of the MW to the LMC. An LMC that maintains a density profile while the center of the MW literally passes right through it is not physical, and sensitivity of the fit mass to the shape of the LMC far out in the dark matter halo is problematic.

\section{Robust Estimation of the LMC mass} \label{sec:LMC_mass}

For the \textit{Erkal2019} potential, all best-fit stream models we found in this paper show approximately the same acceleration from the LMC's gravity due to the LMC at its closest approach to the OCS. It appears that an acceleration of about 1.0--1.3 $\times$ 10$^{8}$ (km/s)$^2$kpc$^{-1}$ is needed to produce the observed bend in the OCS. When the LMC is far away from the OCS, it does not have much of an effect on the system, and certainly does not bend one part of the stream significantly more than another part. We have shown in each optimization that most of the bend in the OCS stream is produced when the LMC is at its closest approach.

In Figure \ref{Fig:all_mass}, we show the enclosed mass as a function of radius for all of the best-fit LMC models in this paper, including all MW potential models. The best-fit Plummer and Hernquist models in Table \ref{tab:best_fit_params} are labeled along with their MW potential names in this figure. Note that enclosed mass calculations are mostly independent of MW potential because they are derived from an analytical equation based on an LMC density profile of choice. Plummer/Hernquist-upper/lower limit models are explained in Section \ref{sec:scale_radius}, and Hernquist-cutoff model is described in Section \ref{sec:LMC_model_cutoff}. This figure shows that all of the models cross each other near 30 kpc, where the 
\begin{longtable*}[t]{ccccccccc}
            \hline
            Model & l & b & d & v$_x$ & v$_y$ & v$_z$ & a$_{LMC}$ & M$_{LMC}$ \\
            & ($^\circ$) & ($^\circ$) & (kpc) & (kpc/Gyr) & (kpc/Gyr) & (kpc/Gyr) & (kpc) & ($10^{10}$M$_\odot$) \\ 
            \hline 
            P-Erkal2019 & 299 & 5.50 & 18.4 & -257.9 & 5.8 & 242.9 & 9.48 & 5.5 \\ 
            P-Salas2019 & 284.10 & 27.782 & 16.8 & -246.715 & 68.751 & 203.042 & 9.48 & 5.5 \\
            P-Mendelsohn2022 & 284.10 & 27.782 & 16.8 &  -228.715 & 72.751 & 174.042 & 9.99 & 6.0 \\
            P-MMWPotential2014 & 299 & 5.75 & 17.8 & -236.4 & -3.4 & 231.1 & 9.48 & 5.5 \\
            H-Erkal2019 & 299 & 5.50 & 18.4 & -257.9 & 4.8 & 243.9 & 11.9 & 9.5 \\ 
            H-Salas2019 & 284.10 & 27.782 & 16.8 & -246.715 & 67.751 & 204.042 & 11.9 & 9.5 \\
            H-Mendelsohn2022 & 284.10 & 27.782 & 16.8 &  -230.715 & 71.751 & 174.042 & 11.9 & 9.5 \\
            H-MMWPotential2014 & 299 & 5.75 & 17.8 & -237.4 & -5.4 & 234.1 & 11.3 & 9.0 \\
            P-upper limit & 299 & 5.50 & 18.4 & -257.9 & 3.8 & 239.9 & 7.47 & 5.5 \\
            P-lower limit & 299 & 5.50 & 18.4 & -257.9 & 5.8 & 241.9 & 13.7 & 6.5 \\
            H-upper limit & 299 & 5.50 & 18.4 & -257.9 & 2.8 & 240.9 & 6.68 & 7.5 \\ 
            H-lower limit & 299 & 5.50 & 18.4 & -256.9 & 5.8 & 244.9 & 32.1 & 22 \\ 
            H-cutoff & 299 & 5.50 & 18.4 & -257.9 & 4.8 & 245.9 & 39.8 & 4.5($<$16.4 kpc) \\ 
            \hline
\caption[table heading]{Best-fit parameters. P: Plummer LMC, H: Hernquist LMC.}
\label{tab:best_fit_params}
\end{longtable*} 
\noindent mean of the mass enclosed is M($<r$) = $4.93 \times 10^{10}$ M$_\odot$. The standard error of the mean is $0.05\times 10^{10}$ M$_\odot$. At large radius from the LMC's center of mass, any LMC mass has negligible effect on the path of the OCS, and therefore the total LMC mass is largely unconstrained. The estimated LMC total mass from fitting the OCS depends strongly on its assumed model because the model determines how much mass is at large distance from the LMC. 

What the OCS can constrain is the LMC mass within 30 kpc. All of the M$_{LMC}$($<$ 30 kpc) mass estimates are in the range 4.7 $\times 10^{10}$ M$_\odot$ $<$ M$_{LMC}$($<$ 30 kpc) $<$ 5.1 $\times 10^{10}$ M$_\odot$.  From the average and the standard deviation of the best-fit masses for each model at 30 kpc, we estimate that the current LMC mass within 30 kpc is $4.93 \pm 0.05 \times 10^{10}$ M$_\odot$. Note that this is an even smaller error than the \cite{Van2014} constraint within 8.7 kpc. The tidal radius of an LMC of this mass is about 16.9 kpc, meaning that our measured mass approximates the current bound mass of the LMC.

\begin{figure}
    \centering
    \includegraphics[width=0.45\textwidth]{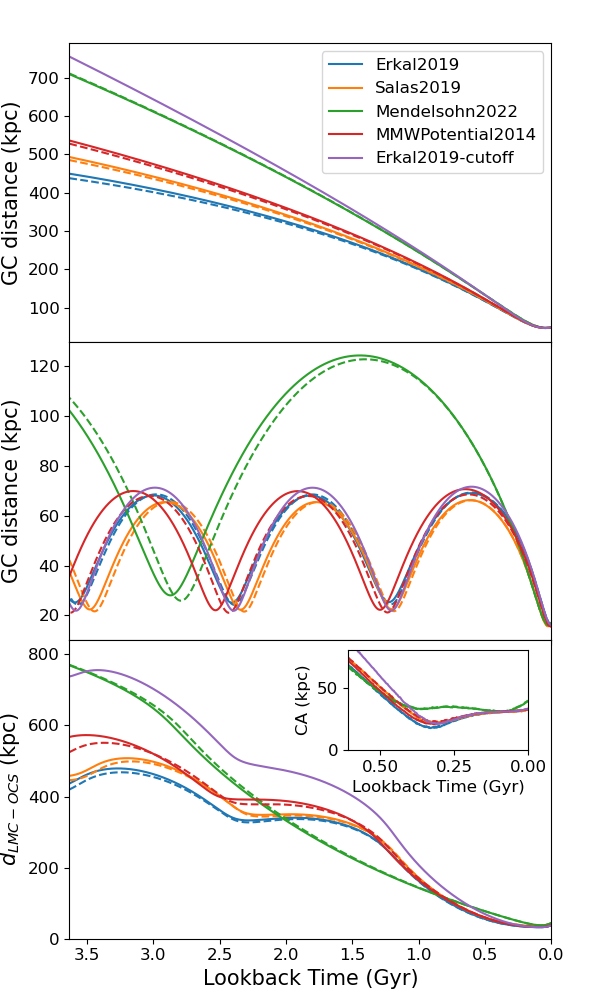}
    \caption{Orbits of the LMC (top), OCS progenitor (middle), and the distance between them (bottom) in each best-fit MW potential case. The inset in the bottom shows the closest approach of the LMC to the OCS. Because the closest approach of the LMC to any point on the OCS is equal to or smaller than the distance between the LMC and the progenitor, the inset curve is slightly less than the corresponding portion of the curve in the lower panel. The solid lines are Hernquist LMC models, and the dashed lines are Plummer LMC models. The orbits made with \textit{Erkal2019} potential are colored blue, the \textit{Salas2019} potential is colored orange, the \textit{Mendelsohn2022} potential is green, the \textit{MMWPotential2014} potential is red, and lastly the cutoff Hernquist LMC model with the \textit{Erkal2019} potential is colored purple. The orbits of both the LMC and OCS with \textit{Mendelsohn2022} are completely different from other MW potential models. This potential model has a smaller rotation curve (see Figure \ref{Fig:MW_potential}), thus a less massive MW. Because the LMC can more easily escape from a less massive MW, the radius of the LMC orbit is larger for the \textit{Mendelsohn2022} case. The radius and period of the OCS orbit also become larger for the same reason. With our adopted values for the current position and velocity of the LMC, the LMC has only recently interacted with the OCS for a short period of time (within the last 500 Myr).}
    \label{Fig:orbit}
\end{figure}

\begin{figure}
    \centering
    \includegraphics[width=0.45\textwidth]{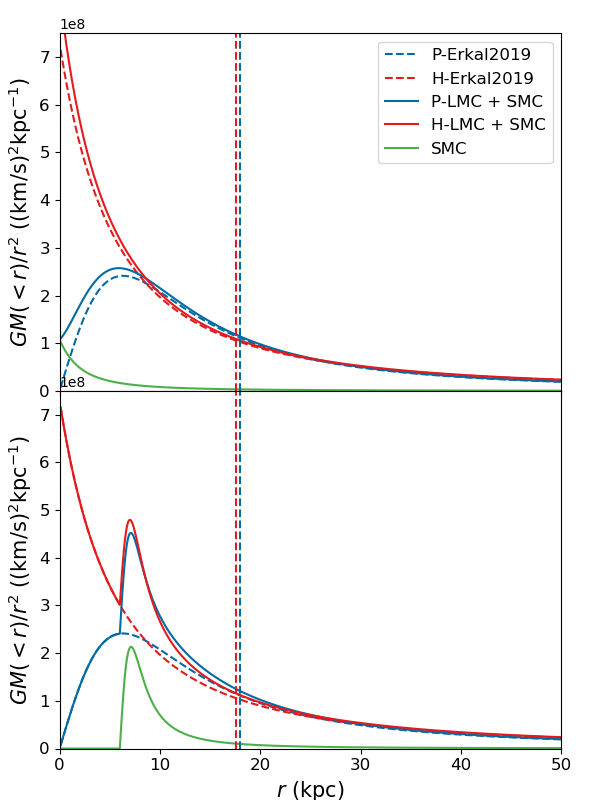}
    \caption{Acceleration of the LMC and SMC pair. Top panel: The dashed lines are our best-fit Plummer LMC (blue) and Hernquist LMC (red) accelerations as a function of distance from the center of the LMC. The solid green line is the acceleration of the SMC only; here the SMC was modeled with a mass of 1.4 $\times$ 10$^{9}$ M$_\odot$ \citep{Teodoro2019} and a scale radius of 1.5895 kpc (the scale radius was selected so that it is likely still bound at the end of the simulation). The vertical lines are the closest approach of the LMC over the course of each simulation. We then added the acceleration of the Plummer SMC to these LMC accelerations (solid lines). Bottom panel: Here, we changed the distance of the SMC model so that the acceleration of the pair at the closest approach would increase to the amount that affects the OCS by the same amount as if we changed the LMC mass by 3.0 $\times$ 10$^{10}$ M$_\odot$ (see Figure \ref{Fig:OCS_fit_Hernquist}). Although the SMC does not affect the OCS much at present day, our analysis suggests that if in the past the SMC was $\backsim$6 kpc closer to the OCS than the LMC was, it would affect the OCS significantly.}
    \label{Fig:SMC_accel}
\end{figure}

\section{Discussion} \label{sec:discussion}

In this study, we estimated that the current mass of the LMC within 30 kpc is $4.93 \times 10^{10}$ M$_\odot$. If the total mass of the best-fit profile is integrated to infinity, we would find total masses anywhere between $4.5 \times 10^{10}$ M$_\odot$ and $2.2 \times 10^{11}$ M$_\odot$ or more, depending on the LMC's radial profile. We have suggested that the OCS orbit is most sensitive to the acceleration from the LMC when it is at closest approach to the stream. 

In Figure \ref{Fig:orbit}, we show the orbits of the LMC (top) and OCS progenitors (middle) that were used in this work. The bottom panel shows their distance from each other, and the inset in this bottom panel shows the closest approach of the LMC to the OCS. Note that the inset curve has a smaller distance than the lower panel because the closest approach to the stream is always the same or smaller than the distance between the LMC and the progenitor. The closest approach under the \textit{Erkal2019}, \textit{Salas2019}, and \textit{MMWPotential2014} potentials is between 17 to 23 kpc at which an acceleration of about 1.0--1.3 $\times$ 10$^{8}$ (km/s)$^2$kpc$^{-1}$ is exerted. The \textit{Mendelsohn2022} case, however, requires an acceleration of 0.5 $\times$ 10$^{8}$ (km/s)$^2$kpc$^{-1}$ at the closest approach of 31 kpc. The required acceleration depends to some extent on the assumed MW potential. 

Figure \ref{Fig:orbit} shows that the LMC has only recently interacted, and for a short period of time. The solid lines are Hernquist LMC models, and the dashed lines are Plummer LMC models. The orbits of both the LMC and OCS with \textit{Mendelsohn2022} are completely different from other MW potential models. As we observed in Figure \ref{Fig:MW_potential}, \textit{Mendelsohn2022} has a smaller rotation curve and therefore a less massive MW model. Since the LMC can more easily escape from a less massive MW, the LMC orbit extends to larger radii. Similarly, the OCS orbit also has a larger radius and period. On the other hand, other potential cases show similar orbits for both the LMC and OCS. Note that in all cases, the LMC has been close enough to significantly interact with the OCS only for the past 500 Myr. Most of the interaction occurs in a 200 Myr time period, which is much less than the orbital period of the OCS.
                            
One shortcoming of our study is that it was assumed that the shape and mass of the LMC do not change over time (in particular it does not get tidally disrupted as it falls into our galaxy during the simulation). Given the orbits of the OCS and LMC, our mass measurement technique finds the current mass of the LMC within 30 kpc. However, the infall mass of the LMC could be more massive than our estimate. \cite{Garavito-Camargo_2021} found that the current bound mass of the LMC is 4--7 $\times$ 10$^{10}$ M$_\odot$ if the initial infall mass is 8--25 $\times$ 10$^{10}$ M$_\odot$. The mass loss was estimated from basis function expansions of N-body snapshots. Note that their bound particles can extend a lot farther (up to $\sim$ 60 kpc) in their elongated LMC shape. Our spherical models have limitations in measuring the total bound mass as we cannot track the LMC's bound particles as the LMC tidally distorts. To the extent that our spherical approximation can be directly compared with their live simulation, our result favors an infall initial mass at the low end of their range, closer to 1.0 $\times$ 10$^{11}$ M$_\odot$. 

Another shortcoming of our spherical model is that it does not include the time-dependent gravitational effects of the SMC, or alternatively we assumed the effect of the SMC was included in the spherical simulation of the LMC. This is plausibly justified because the acceleration from the SMC at the present time is much smaller than that of the LMC. In the top panel in Figure \ref{Fig:SMC_accel}, we show the acceleration of the LMC and SMC pair at present time. The origin is the center of the LMC, and the x-axis is the direction from the center of the SMC to the closest part of the stream at present time. The dashed lines are our best-fit Plummer LMC (blue) and Hernquist LMC (red) accelerations. The dashed vertical lines are the closest approach of the LMC over the course of each simulation. We then added the acceleration of the Plummer SMC with a mass of 1.4 $\times$ 10$^{9}$ M$_\odot$ \citep{Teodoro2019} and a scale radius of 1.5895 kpc (the scale radius was selected so that it is likely still bound at the end of the simulation) to these LMC accelerations (solid blue and red lines). The solid green line is the acceleration of the SMC only. At present time, we do not expect the SMC to affect the OCS much as there is almost no acceleration change after adding the SMC. 

Nevertheless, it is possible that the SMC was closer to the OCS than the LMC in the past 0.5 Gyr \citep{Patel2020}. Further in the past, the LMC is farther from the MW than the OCS orbit (see Figure \ref{Fig:orbit}) and would not be expected to produce enough acceleration to displace the OCS stream. In the bottom panel in Figure \ref{Fig:SMC_accel}, we changed the distance of the SMC model so that the acceleration of the pair at the closest approach would increase to the amount that affects the OCS by M$_{LMC}$ = 3.0 $\times$ 10$^{10}$ M$_\odot$ (see Figure \ref{Fig:OCS_fit_Hernquist}). This plot shows that if in the past 0.5 Gyr the SMC came $ \backsim$6 kpc closer than the LMC's distance from the OCS, then the SMC could have caused the OCS to change its path. However, the orbit of the SMC is poorly known \citep{D'Onghia2016}. We considered including the SMC, but were not able to produce any substantive results because of the poorly known orbit.

The SMC is not the only halo object that could affect the path of the OCS. A massive merger event is known to cause the disk to tilt \citep{Huang1997}. \cite{Nibauer2024} found that the titling disk can make the streams more diffuse or narrow, depending on the orbit inclination and the tilting disk direction. Moreover, \cite{Lilleengen2023} showed that the OCS is affected by the deformation of the dark matter halos of the LMC and MW. Therefore, the tilting disk and the shape of the LMC and the MW halo can affect the LMC mass estimates and should be investigated in future work.

\section{Conclusions} \label{sec:conclusion}

In this paper, we measured the LMC mass from its effect on the path of the OCS and explored potential systematic errors in the estimates. We summarize our findings as follows:

\begin{enumerate}

  \item All of the best-fit LMC models in this paper showed approximately the same mass within 30 kpc, which is M$_{LMC}(<r) = $ 4.93 $\pm$ 0.05 $\times 10^{10}$ M$_\odot$ with a range of 4.7 $\times 10^{10}$ M$_\odot$ $<$ M$_{LMC}$($<$ 30 kpc) $<$ 5.1 $\times 10^{10}$ M$_\odot$. Since the tidal radius of a mass of this size in a MW potential is $ \backsim$16.9 kpc, we are approximately measuring the current bound mass of the LMC. For three of our MW models, the LMC exerted an acceleration of 1.0--1.3 $\times$ 10$^{8}$ (km/s)$^2$kpc$^{-1}$ at closest approach to the OCS; the closest approach was 17--23 kpc. For the lower mass \textit{Mendelsohn2022} MW potential, an acceleration of 0.5 $\times$ 10$^{8}$ (km/s)$^2$kpc$^{-1}$ was exerted at closest approach of 31 kpc; with this potential we also struggled to match the path of the OCS. The interaction occurred within the last 500 Myr and lasted for about 200 Myr. 

  \item The total LMC mass depends not only on the mass within 30 kpc, but also on all of the mass outside this radius. By choosing radial profiles that put large amounts of mass at large radii, the presumed total mass of the LMC can be 2.2 $\times 10^{11}$ M$_\odot$ or even higher. Modeling the LMC with a Plummer profile produced a total LMC mass estimate of M$_{LMC} = $ 5.5--6.0 $\times 10^{10}$ M$_\odot$. Modeling the LMC as the Hernquist profile produced a mass estimate of M$_{LMC} = $ 9.0--9.5 $\times 10^{10}$ M$_\odot$. These results show that by switching the LMC model, the total mass of the LMC changes by 3.5--4.0 $\times 10^{10}$ M$_\odot$. The selection of the scale radius of the LMC density profile also influences the path of the OCS. Using the allowed range of the scale radius, the mass of the LMC can change by 16.5 $\times 10^{10}$ M$_\odot$ or more. Modeling the LMC with a truncated Hernquist profile at its tidal radius produced an estimate of M$_{LMC} = $ 4.5 $\times 10^{10}$ M$_\odot$. These numbers show that fitting the OCS does not constrain the radial profile of the LMC, and does not predict the total initial or current mass at large radii.  

  \item The assumed MW potential has little effect on the LMC mass estimate from the path of the OCS stream, as long as the potential is consistent with the observed rotation curve.
   
  \item There is a difference in the path of the OCS between N-body simulations and traditional particle-spray modeling; the stream modeled with N-body simulation is wider and slightly tilted from the stream modeled with particle-spray. The difference between the two simulations arises from the fact that the N-body particles are stripped at a much larger range of radii, and therefore produce a wider stream. Having a wide range of energies also results in the center of the leading tidal stream being pulled closer to the Galactic center and the trailing tidal stream being pushed farther from the Galactic center. This causes the N-body simulation stream to be tilted from the particle-spray stream. This difference between the two simulation techniques changes the estimated total mass of the LMC by $3.0 \times 10^{10}$ M$_\odot$.

  \item The LMC mass estimate is not affected by whether the OCS progenitor is modeled as a single-component or two-component (baryon and dark matter) Plummer sphere. While the number of components affects distribution of stars along the stream and has a small effect on the stream width, it does not significantly change the path of the stream.
  
\end{enumerate} 

While our measurements do not measure the initial mass of the LMC before infall, comparison of our present LMC mass with the simulations of \cite{Garavito-Camargo_2021} suggests an initial mass of $\backsim1\times 10^{11}$ M$_\odot$ is reasonable.


\begin{acknowledgments}

We thank A. Price-Whelan, N. Garavito-Camargo, K. V. Johnston, and M. Lisantifor for valuable discussions and comments. We also thank the Nearby Universe group at the Center for Computational Astrophysics (CCA) for insightful comments and Lee Newberg for careful proofreading. This work was supported by NSF grant 24-06594. H. T. Warren was partially supported by a NASA/NY Space Grant fellowship. H. T. Warren and H. J. Newberg thank the CCA at Flatiron Institute for hospitality while (a portion of) this research was carried out; the Flatiron Institute is division of the Simons Foundation.

This work made use of galpy: \url{https://github.com/jobovy/galpy} \citep{Bovy2015} and Astropy: \url{http://www.astropy.org} a community-developed core Python package and an ecosystem of tools and resources for astronomy \citep{astropy:2013, astropy:2018, astropy:2022}.

\end{acknowledgments}

\appendix

Because our N-body simulations of DG tidal disruption are intended to be used for optimizations, we do not evolve them in isolation prior to the main simulation. Here we show that the generated DGs are stable. We use exactly the same DG parameters for the stability calculations as were used to generate the DGs in each simulation. 

To assess the stability of DGs in a null potential, we use the Kullback-Leibler (KL) divergence as a way to measure how much the simulated probability distribution diverges from the theoretical probability distribution of particle positions over time. The KL divergence is calculated using the following equation \citep{Kullback-Leibler1951}:
\begin{equation}
    D_{KL} (P || Q) = \sum_{x \in \chi} P(x) \ln \frac{P(x)}{Q(x)},
\end{equation}
where $P$ is the theoretical probability distribution and $Q$ is the simulated probability distribution. A KL divergence closer to 0 indicates greater similarity between the two probability distributions and, when calculated over time, gives a metric for the stability of the simulated DG. 

\begin{figure}
    \centering
    \includegraphics[width=.45\textwidth]{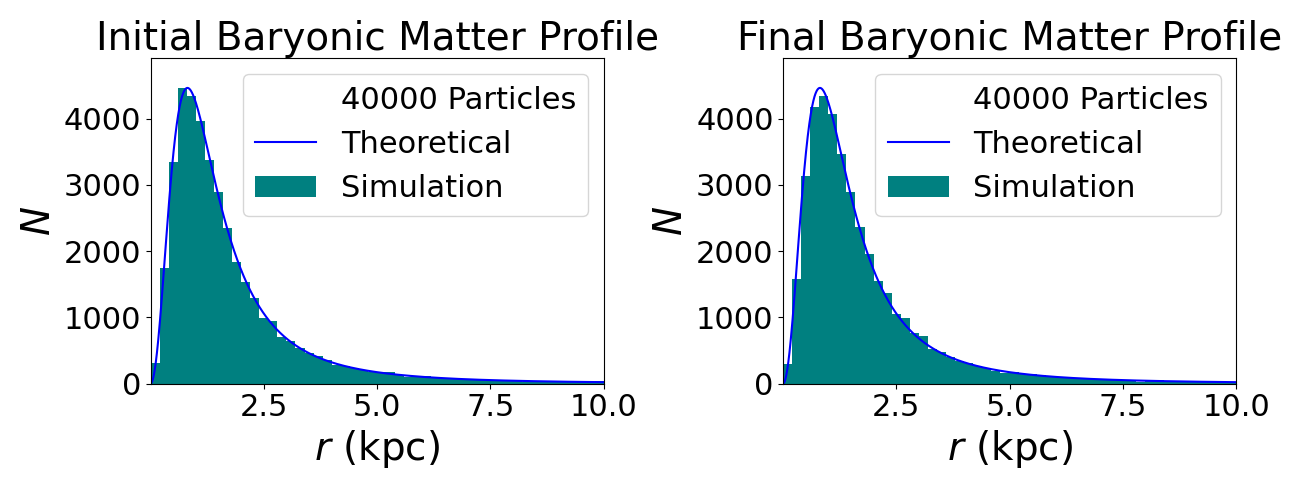}
    \caption{The histograms show the initial and final number density distribution from the simulated particles of the single component Plummer DG. The figure shows clear agreement between the simulated data in teal and the expected number distribution in blue even after evolving for 4 Gyrs in a null potential.}
    \label{Fig:single_plummer_hist}
\end{figure}

\begin{figure}
    \centering
    \includegraphics[width=.45\textwidth]{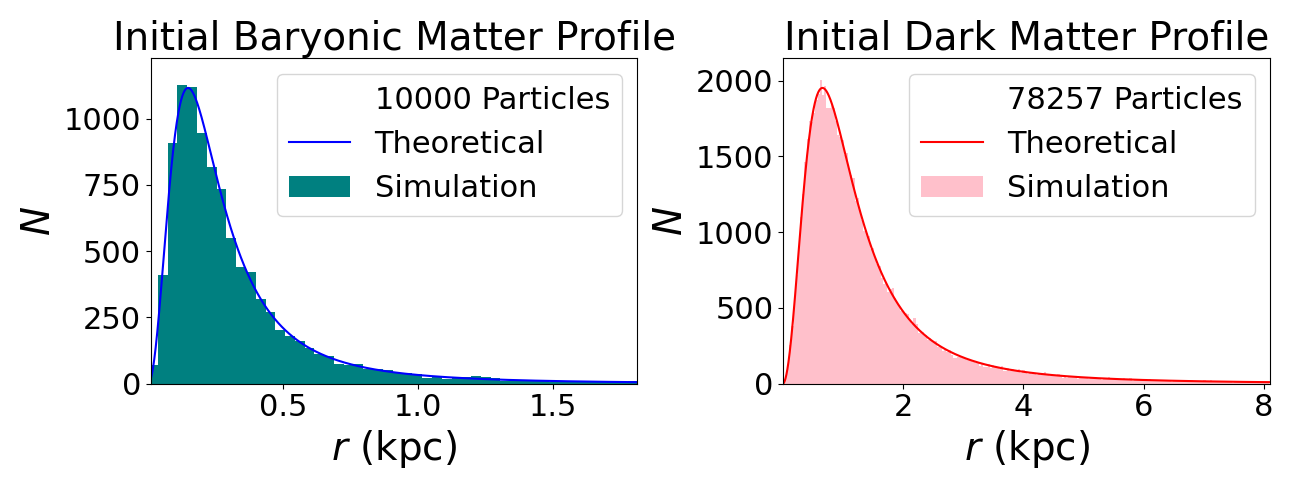}
    \includegraphics[width=.45\textwidth]{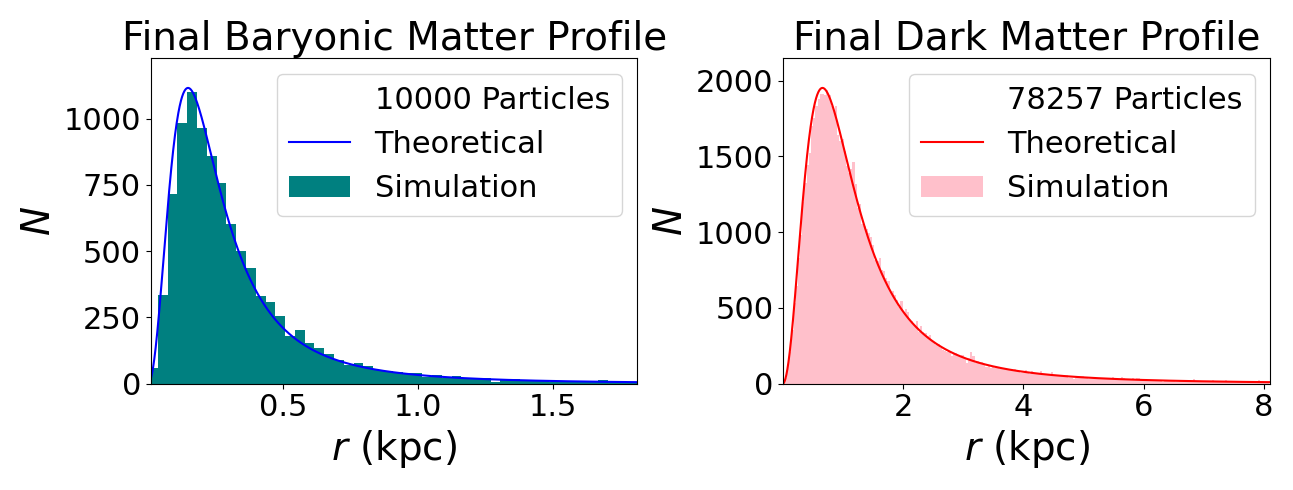}
    \caption{The histograms for the initial and final number density distribution for the two component Plummer-Plummer DG is shown. Like the single Plummer model, both the baryons (left) and the dark matter (right) agree with the expected number distribution even after evolving for 4 Gyrs.}
    \label{Fig:double_plummer_hist}
\end{figure}

\begin{figure}
    \centering
    \includegraphics[width=.45\textwidth]{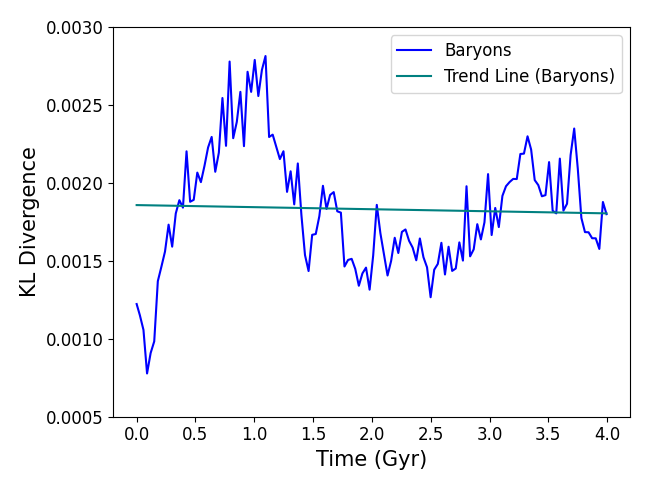}
    \caption{The KL divergence for a single-component Plummer DG, calculated every tenth timestep (3.05 million years) over 4 Gyr shown in blue. The trend line (teal) indicates high stability, with fluctuations in KL divergence limited to approximately 0.002.}
   \label{Fig:kl_divergence_single_plummer}
\end{figure}

\begin{figure}
    \centering
    \includegraphics[width=.45\textwidth]{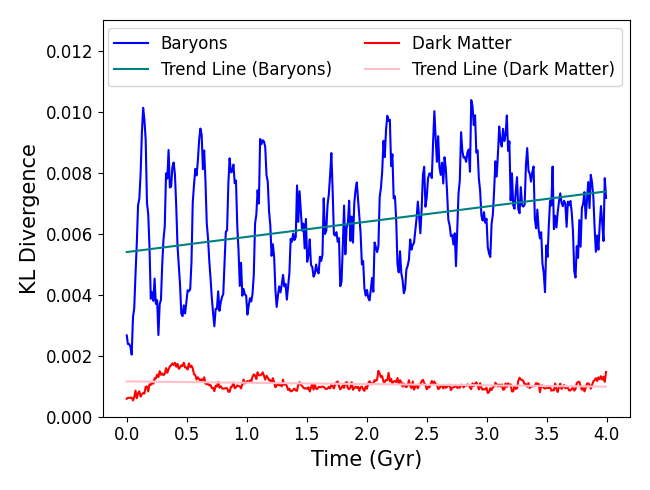}
    \caption{The KL divergence for a two-component Plummer DG for both the baryon density profile (blue) and the dark matter density profile (red), calculated every tenth timestep (1.06 million years) over 4 Gyr. The dark matter shows very little fluctuations in comparison to the baryons. This is expected since the number of dark matter particles is much greater than the number of baryons. The slight trend towards instability seen in teal and the sinusoidal fluctuations of the KL divergence for the baryons is due to the fact that the baryon particle mass is ten percent the mass of the dark matter particles. This induces a small amount of dynamical friction in the center of the DG that causes a very slight trend towards instability.}  
   \label{Fig:kl_divergence_double_plummer}
\end{figure}

To calculate the KL divergence as a function of time, we evolve the simulated DG in a null potential for 4 Gyr. At each timestep, the particles are binned as shown in Figures \ref{Fig:single_plummer_hist} and \ref{Fig:double_plummer_hist}. Each bin height, $Q^*(x)$ with $*$ denoting the non-normalized count, is the number of particles within a shell of radius $r(x)$ and bin width $dr$. The corresponding theoretical distribution is:
\begin{equation}
P^*(x) = dN(x) = \frac{4 \pi r^2 \rho(r(x))}{m_{particle}} dr,
\end{equation}
where $\rho(r(x))$ is the density in bin $x$ at radius $r(x)$ and $m_{particle}$ is the mass per particle. Note that we only consider DGs that are spherically symmetric. 

We calculate the KL divergence within a radius range from the second bin to 4 times the scale radius, where most of the particles are concentrated. To avoid a bin height of 0, all bins are increased by $10^{-6}$ counts, and the counts within the selected range are normalized to obtain the probability distribution functions, $P(x)$ and $Q(x)$, as a function of bin within the selected range. Figures \ref{Fig:kl_divergence_single_plummer} and \ref{Fig:kl_divergence_double_plummer} show the KL divergence every tenth timestep over the course of the 4 Gyr simulation for the single-component and double-component Plummer models, respectively.  

The initial and final number density profiles for both the single-component and double-component Plummer profiles are shown in Figures \ref{Fig:single_plummer_hist} and \ref{Fig:double_plummer_hist}, respectively. Both figures show that the simulated particle density distributions match the theoretical density distributions in its initial state and after a 4 Gyr simulation in a null potential. 

This stability is corroborated by the KL divergence values in Figures \ref{Fig:kl_divergence_single_plummer} and \ref{Fig:kl_divergence_double_plummer}, which exhibit minimal fluctuations over the simulation period. Typically, we consider the DG to be unstable if the KL divergence rises to greater than three times its initial value. Instability is usually observed within the first 500 million years, where the KL divergence over time is sinusoidal with an upward trend. Note that in the case of the double-Plummer profile, the baryon component has minor oscillations and increases slightly due to DF since the baryon particles are a tenth the mass of the dark matter particles. When the mass per particle of both components are the same, the oscillatory behavior and upward trend are absent.

\bibliographystyle{aasjournal}
\bibliography{references.bib}

\begin{thebibliography}{}
\expandafter\ifx\csname natexlab\endcsname\relax\def\natexlab#1{#1}\fi
\providecommand{\url}[1]{\href{#1}{#1}}
\providecommand{\dodoi}[1]{doi:~\href{http://doi.org/#1}{\nolinkurl{#1}}}
\providecommand{\doeprint}[1]{\href{http://ascl.net/#1}{\nolinkurl{http://ascl.net/#1}}}
\providecommand{\doarXiv}[1]{\href{https://arxiv.org/abs/#1}{\nolinkurl{https://arxiv.org/abs/#1}}}

\bibitem[{{Aarseth} {et~al.}(1974){Aarseth}, {Henon}, \& {Wielen}}]{Aarseth1974}
{Aarseth}, S.~J., {Henon}, M., \& {Wielen}, R. 1974, \aap, 37, 183

\bibitem[{{Astropy Collaboration} {et~al.}(2013){Astropy Collaboration}, {Robitaille}, {Tollerud}, {Greenfield}, {Droettboom}, {Bray}, {Aldcroft}, {Davis}, {Ginsburg}, {Price-Whelan}, {Kerzendorf}, {Conley}, {Crighton}, {Barbary}, {Muna}, {Ferguson}, {Grollier}, {Parikh}, {Nair}, {Unther}, {Deil}, {Woillez}, {Conseil}, {Kramer}, {Turner}, {Singer}, {Fox}, {Weaver}, {Zabalza}, {Edwards}, {Azalee Bostroem}, {Burke}, {Casey}, {Crawford}, {Dencheva}, {Ely}, {Jenness}, {Labrie}, {Lim}, {Pierfederici}, {Pontzen}, {Ptak}, {Refsdal}, {Servillat}, \& {Streicher}}]{astropy:2013}
{Astropy Collaboration}, {Robitaille}, T.~P., {Tollerud}, E.~J., {et~al.} 2013, \aap, 558, A33, \dodoi{10.1051/0004-6361/201322068}

\bibitem[{{Astropy Collaboration} {et~al.}(2018){Astropy Collaboration}, {Price-Whelan}, {Sip{\H{o}}cz}, {G{\"u}nther}, {Lim}, {Crawford}, {Conseil}, {Shupe}, {Craig}, {Dencheva}, {Ginsburg}, {Vand erPlas}, {Bradley}, {P{\'e}rez-Su{\'a}rez}, {de Val-Borro}, {Aldcroft}, {Cruz}, {Robitaille}, {Tollerud}, {Ardelean}, {Babej}, {Bach}, {Bachetti}, {Bakanov}, {Bamford}, {Barentsen}, {Barmby}, {Baumbach}, {Berry}, {Biscani}, {Boquien}, {Bostroem}, {Bouma}, {Brammer}, {Bray}, {Breytenbach}, {Buddelmeijer}, {Burke}, {Calderone}, {Cano Rodr{\'\i}guez}, {Cara}, {Cardoso}, {Cheedella}, {Copin}, {Corrales}, {Crichton}, {D'Avella}, {Deil}, {Depagne}, {Dietrich}, {Donath}, {Droettboom}, {Earl}, {Erben}, {Fabbro}, {Ferreira}, {Finethy}, {Fox}, {Garrison}, {Gibbons}, {Goldstein}, {Gommers}, {Greco}, {Greenfield}, {Groener}, {Grollier}, {Hagen}, {Hirst}, {Homeier}, {Horton}, {Hosseinzadeh}, {Hu}, {Hunkeler}, {Ivezi{\'c}}, {Jain}, {Jenness}, {Kanarek}, {Kendrew}, {Kern}, {Kerzendorf}, {Khvalko}, {King}, {Kirkby}, {Kulkarni},
  {Kumar}, {Lee}, {Lenz}, {Littlefair}, {Ma}, {Macleod}, {Mastropietro}, {McCully}, {Montagnac}, {Morris}, {Mueller}, {Mumford}, {Muna}, {Murphy}, {Nelson}, {Nguyen}, {Ninan}, {N{\"o}the}, {Ogaz}, {Oh}, {Parejko}, {Parley}, {Pascual}, {Patil}, {Patil}, {Plunkett}, {Prochaska}, {Rastogi}, {Reddy Janga}, {Sabater}, {Sakurikar}, {Seifert}, {Sherbert}, {Sherwood-Taylor}, {Shih}, {Sick}, {Silbiger}, {Singanamalla}, {Singer}, {Sladen}, {Sooley}, {Sornarajah}, {Streicher}, {Teuben}, {Thomas}, {Tremblay}, {Turner}, {Terr{\'o}n}, {van Kerkwijk}, {de la Vega}, {Watkins}, {Weaver}, {Whitmore}, {Woillez}, {Zabalza}, \& {Astropy Contributors}}]{astropy:2018}
{Astropy Collaboration}, {Price-Whelan}, A.~M., {Sip{\H{o}}cz}, B.~M., {et~al.} 2018, \aj, 156, 123, \dodoi{10.3847/1538-3881/aabc4f}

\bibitem[{{Astropy Collaboration} {et~al.}(2022){Astropy Collaboration}, {Price-Whelan}, {Lim}, {Earl}, {Starkman}, {Bradley}, {Shupe}, {Patil}, {Corrales}, {Brasseur}, {N{"o}the}, {Donath}, {Tollerud}, {Morris}, {Ginsburg}, {Vaher}, {Weaver}, {Tocknell}, {Jamieson}, {van Kerkwijk}, {Robitaille}, {Merry}, {Bachetti}, {G{"u}nther}, {Aldcroft}, {Alvarado-Montes}, {Archibald}, {B{'o}di}, {Bapat}, {Barentsen}, {Baz{'a}n}, {Biswas}, {Boquien}, {Burke}, {Cara}, {Cara}, {Conroy}, {Conseil}, {Craig}, {Cross}, {Cruz}, {D'Eugenio}, {Dencheva}, {Devillepoix}, {Dietrich}, {Eigenbrot}, {Erben}, {Ferreira}, {Foreman-Mackey}, {Fox}, {Freij}, {Garg}, {Geda}, {Glattly}, {Gondhalekar}, {Gordon}, {Grant}, {Greenfield}, {Groener}, {Guest}, {Gurovich}, {Handberg}, {Hart}, {Hatfield-Dodds}, {Homeier}, {Hosseinzadeh}, {Jenness}, {Jones}, {Joseph}, {Kalmbach}, {Karamehmetoglu}, {Ka{l}uszy{'n}ski}, {Kelley}, {Kern}, {Kerzendorf}, {Koch}, {Kulumani}, {Lee}, {Ly}, {Ma}, {MacBride}, {Maljaars}, {Muna}, {Murphy}, {Norman}, {O'Steen},
  {Oman}, {Pacifici}, {Pascual}, {Pascual-Granado}, {Patil}, {Perren}, {Pickering}, {Rastogi}, {Roulston}, {Ryan}, {Rykoff}, {Sabater}, {Sakurikar}, {Salgado}, {Sanghi}, {Saunders}, {Savchenko}, {Schwardt}, {Seifert-Eckert}, {Shih}, {Jain}, {Shukla}, {Sick}, {Simpson}, {Singanamalla}, {Singer}, {Singhal}, {Sinha}, {Sip{H{o}}cz}, {Spitler}, {Stansby}, {Streicher}, {{{S}}umak}, {Swinbank}, {Taranu}, {Tewary}, {Tremblay}, {Val-Borro}, {Van Kooten}, {Vasovi{'c}}, {Verma}, {de Miranda Cardoso}, {Williams}, {Wilson}, {Winkel}, {Wood-Vasey}, {Xue}, {Yoachim}, {Zhang}, {Zonca}, \& {Astropy Project Contributors}}]{astropy:2022}
{Astropy Collaboration}, {Price-Whelan}, A.~M., {Lim}, P.~L., {et~al.} 2022, \apj, 935, 167, \dodoi{10.3847/1538-4357/ac7c74}

\bibitem[{{Barnes} \& {Hut}(1986)}]{Barnes1986}
{Barnes}, J., \& {Hut}, P. 1986, \nat, 324, 446, \dodoi{10.1038/324446a0}

\bibitem[{Barnes(2001)}]{barnes2001}
Barnes, J.~E. 2001, Treecode guide,  February

\bibitem[{{Besla} {et~al.}(2007){Besla}, {Kallivayalil}, {Hernquist}, {Robertson}, {Cox}, {van der Marel}, \& {Alcock}}]{Besla2007}
{Besla}, G., {Kallivayalil}, N., {Hernquist}, L., {et~al.} 2007, \apj, 668, 949, \dodoi{10.1086/521385}

\bibitem[{{Besla} {et~al.}(2012){Besla}, {Kallivayalil}, {Hernquist}, {van der Marel}, {Cox}, \& {Kere{\v{s}}}}]{Besla2012}
---. 2012, \mnras, 421, 2109, \dodoi{10.1111/j.1365-2966.2012.20466.x}

\bibitem[{{Bonaca} {et~al.}(2019){Bonaca}, {Hogg}, {Price-Whelan}, \& {Conroy}}]{Bonaca2019}
{Bonaca}, A., {Hogg}, D.~W., {Price-Whelan}, A.~M., \& {Conroy}, C. 2019, \apj, 880, 38, \dodoi{10.3847/1538-4357/ab2873}

\bibitem[{{Bovy}(2015)}]{Bovy2015}
{Bovy}, J. 2015, \apjs, 216, 29, \dodoi{10.1088/0067-0049/216/2/29}

\bibitem[{{Burke}(1957)}]{Burke1957}
{Burke}, B.~F. 1957, \aj, 62, 90, \dodoi{10.1086/107463}

\bibitem[{{Chandrasekhar}(1943)}]{Chandrasekhar1943}
{Chandrasekhar}, S. 1943, \apj, 97, 255, \dodoi{10.1086/144517}

\bibitem[{{Chen} {et~al.}(2024){Chen}, {Valluri}, {Gnedin}, \& {Ash}}]{Chen2024}
{Chen}, Y., {Valluri}, M., {Gnedin}, O.~Y., \& {Ash}, N. 2024, arXiv e-prints, arXiv:2408.01496, \dodoi{10.48550/arXiv.2408.01496}

\bibitem[{{Correa Magnus} \& {Vasiliev}(2022)}]{Magnus2022}
{Correa Magnus}, L., \& {Vasiliev}, E. 2022, \mnras, 511, 2610, \dodoi{10.1093/mnras/stab3726}

\bibitem[{{de Salas} {et~al.}(2019){de Salas}, {Malhan}, {Freese}, {Hattori}, \& {Valluri}}]{Salas2019}
{de Salas}, P.~F., {Malhan}, K., {Freese}, K., {Hattori}, K., \& {Valluri}, M. 2019, \jcap, 2019, 037, \dodoi{10.1088/1475-7516/2019/10/037}

\bibitem[{{Deason} {et~al.}(2015){Deason}, {Wetzel}, {Garrison-Kimmel}, \& {Belokurov}}]{Deason2015}
{Deason}, A.~J., {Wetzel}, A.~R., {Garrison-Kimmel}, S., \& {Belokurov}, V. 2015, \mnras, 453, 3568, \dodoi{10.1093/mnras/stv1939}

\bibitem[{{Di Teodoro} {et~al.}(2019){Di Teodoro}, {McClure-Griffiths}, {Jameson}, {D{\'e}nes}, {Dickey}, {Stanimirovi{\'c}}, {Staveley-Smith}, {Anderson}, {Bunton}, {Chippendale}, {Lee-Waddell}, {MacLeod}, \& {Voronkov}}]{Teodoro2019}
{Di Teodoro}, E.~M., {McClure-Griffiths}, N.~M., {Jameson}, K.~E., {et~al.} 2019, \mnras, 483, 392, \dodoi{10.1093/mnras/sty3095}

\bibitem[{{Diaz} \& {Bekki}(2012)}]{Diaz2012}
{Diaz}, J.~D., \& {Bekki}, K. 2012, \apj, 750, 36, \dodoi{10.1088/0004-637X/750/1/36}

\bibitem[{{D'Onghia} \& {Fox}(2016)}]{D'Onghia2016}
{D'Onghia}, E., \& {Fox}, A.~J. 2016, \araa, 54, 363, \dodoi{10.1146/annurev-astro-081915-023251}

\bibitem[{{Eilers} {et~al.}(2019){Eilers}, {Hogg}, {Rix}, \& {Ness}}]{Eilers2019}
{Eilers}, A.-C., {Hogg}, D.~W., {Rix}, H.-W., \& {Ness}, M.~K. 2019, \apj, 871, 120, \dodoi{10.3847/1538-4357/aaf648}

\bibitem[{{Erkal} \& {Belokurov}(2020)}]{Erkal2020}
{Erkal}, D., \& {Belokurov}, V.~A. 2020, \mnras, 495, 2554, \dodoi{10.1093/mnras/staa1238}

\bibitem[{{Erkal} {et~al.}(2018){Erkal}, {Li}, {Koposov}, {Belokurov}, {Balbinot}, {Bechtol}, {Buncher}, {Drlica-Wagner}, {Kuehn}, {Marshall}, {Mart{\'\i}nez-V{\'a}zquez}, {Pace}, {Shipp}, {Simon}, {Stringer}, {Vivas}, {Wechsler}, {Yanny}, {Abdalla}, {Allam}, {Annis}, {Avila}, {Bertin}, {Brooks}, {Buckley-Geer}, {Burke}, {Carnero Rosell}, {Carrasco Kind}, {Carretero}, {D'Andrea}, {da Costa}, {Davis}, {De Vicente}, {Doel}, {Eifler}, {Evrard}, {Flaugher}, {Frieman}, {Garc{\'\i}a-Bellido}, {Gaztanaga}, {Gerdes}, {Gruen}, {Gruendl}, {Gschwend}, {Gutierrez}, {Hartley}, {Hollowood}, {Honscheid}, {James}, {Krause}, {Maia}, {March}, {Menanteau}, {Miquel}, {Ogando}, {Plazas}, {Sanchez}, {Santiago}, {Scarpine}, {Schindler}, {Sevilla-Noarbe}, {Smith}, {Smith}, {Soares-Santos}, {Sobreira}, {Suchyta}, {Swanson}, {Tarle}, {Tucker}, \& {Walker}}]{Erkal2018}
{Erkal}, D., {Li}, T.~S., {Koposov}, S.~E., {et~al.} 2018, \mnras, 481, 3148, \dodoi{10.1093/mnras/sty2518}

\bibitem[{{Erkal} {et~al.}(2019){Erkal}, {Belokurov}, {Laporte}, {Koposov}, {Li}, {Grillmair}, {Kallivayalil}, {Price-Whelan}, {Evans}, {Hawkins}, {Hendel}, {Mateu}, {Navarro}, {del Pino}, {Slater}, {Sohn}, \& {Orphan Aspen Treasury Collaboration}}]{Erkal2019}
{Erkal}, D., {Belokurov}, V., {Laporte}, C.~F.~P., {et~al.} 2019, \mnras, 487, 2685, \dodoi{10.1093/mnras/stz1371}

\bibitem[{{Fardal} {et~al.}(2015){Fardal}, {Huang}, \& {Weinberg}}]{Fardal2015}
{Fardal}, M.~A., {Huang}, S., \& {Weinberg}, M.~D. 2015, \mnras, 452, 301, \dodoi{10.1093/mnras/stv1198}

\bibitem[{{Fardal} {et~al.}(2019){Fardal}, {van der Marel}, {Law}, {Sohn}, {Sesar}, {Hernitschek}, \& {Rix}}]{Fardal2019}
{Fardal}, M.~A., {van der Marel}, R.~P., {Law}, D.~R., {et~al.} 2019, \mnras, 483, 4724, \dodoi{10.1093/mnras/sty3428}

\bibitem[{{Gaia Collaboration} {et~al.}(2018){Gaia Collaboration}, {Spoto}, {Tanga}, {Mignard}, {Berthier}, {Carry}, {Cellino}, {Dell'Oro}, {Hestroffer}, {Muinonen}, {Pauwels}, {Petit}, {David}, {De Angeli}, {Delbo}, {Fr{\'e}zouls}, {Galluccio}, {Granvik}, {Guiraud}, {Hern{\'a}ndez}, {Ord{\'e}novic}, {Portell}, {Poujoulet}, {Thuillot}, {Walmsley}, {Brown}, {Vallenari}, {Prusti}, {de Bruijne}, {Babusiaux}, {Bailer-Jones}, {Biermann}, {Evans}, {Eyer}, {Jansen}, {Jordi}, {Klioner}, {Lammers}, {Lindegren}, {Luri}, {Panem}, {Pourbaix}, {Randich}, {Sartoretti}, {Siddiqui}, {Soubiran}, {van Leeuwen}, {Walton}, {Arenou}, {Bastian}, {Cropper}, {Drimmel}, {Katz}, {Lattanzi}, {Bakker}, {Cacciari}, {Casta{\~n}eda}, {Chaoul}, {Cheek}, {Fabricius}, {Guerra}, {Holl}, {Masana}, {Messineo}, {Mowlavi}, {Nienartowicz}, {Panuzzo}, {Riello}, {Seabroke}, {Th{\'e}venin}, {Gracia-Abril}, {Comoretto}, {Garcia-Reinaldos}, {Teyssier}, {Altmann}, {Andrae}, {Audard}, {Bellas-Velidis}, {Benson}, {Blomme}, {Burgess}, {Busso}, {Clementini},
  {Clotet}, {Creevey}, {Davidson}, {De Ridder}, {Delchambre}, {Ducourant}, {Fern{\'a}ndez-Hern{\'a}ndez}, {Fouesneau}, {Fr{\'e}mat}, {Garc{\'\i}a-Torres}, {Gonz{\'a}lez-N{\'u}{\~n}ez}, {Gonz{\'a}lez-Vidal}, {Gosset}, {Guy}, {Halbwachs}, {Hambly}, {Harrison}, {Hodgkin}, {Hutton}, {Jasniewicz}, {Jean-Antoine-Piccolo}, {Jordan}, {Korn}, {Krone-Martins}, {Lanzafame}, {Lebzelter}, {L{\"o}}, {Manteiga}, {Marrese}, {Mart{\'\i}n-Fleitas}, {Moitinho}, {Mora}, {Osinde}, {Pancino}, {Recio-Blanco}, {Richards}, {Rimoldini}, {Robin}, {Sarro}, {Siopis}, {Smith}, {Sozzetti}, {S{\"u}veges}, {Torra}, {van Reeven}, {Abbas}, {Abreu Aramburu}, {Accart}, {Aerts}, {Altavilla}, {{\'A}lvarez}, {Alvarez}, {Alves}, {Anderson}, {Andrei}, {Anglada Varela}, {Antiche}, {Antoja}, {Arcay}, {Astraatmadja}, {Bach}, {Baker}, {Balaguer-N{\'u}{\~n}ez}, {Balm}, {Barache}, {Barata}, {Barbato}, {Barblan}, {Barklem}, {Barrado}, {Barros}, {Barstow}, {Bartholom{\'e} Mu{\~n}oz}, {Bassilana}, {Becciani}, {Bellazzini}, {Berihuete}, {Bertone}, {Bianchi},
  {Bienaym{\'e}}, {Blanco-Cuaresma}, {Boch}, {Boeche}, {Bombrun}, {Borrachero}, {Bossini}, {Bouquillon}, {Bourda}, {Bragaglia}, {Bramante}, {Breddels}, {Bressan}, {Brouillet}, {Br{\"u}semeister}, {Brugaletta}, {Bucciarelli}, {Burlacu}, {Busonero}, {Butkevich}, {Buzzi}, {Caffau}, {Cancelliere}, {Cannizzaro}, {Cantat-Gaudin}, {Carballo}, {Carlucci}, {Carrasco}, {Casamiquela}, {Castellani}, {Castro-Ginard}, {Charlot}, {Chemin}, {Chiavassa}, {Cocozza}, {Costigan}, {Cowell}, \& {Crifo}}]{Gaia_Collaboration2018}
{Gaia Collaboration}, {Spoto}, F., {Tanga}, P., {et~al.} 2018, \aap, 616, A13, \dodoi{10.1051/0004-6361/201832900}

\bibitem[{{Garavito-Camargo} {et~al.}(2021){Garavito-Camargo}, {Besla}, {Laporte}, {Price-Whelan}, {Cunningham}, {Johnston}, {Weinberg}, \& {G{\'o}mez}}]{Garavito-Camargo_2021}
{Garavito-Camargo}, N., {Besla}, G., {Laporte}, C. F.~P., {et~al.} 2021, \apj, 919, 109, \dodoi{10.3847/1538-4357/ac0b44}

\bibitem[{{Gibbons} {et~al.}(2014){Gibbons}, {Belokurov}, \& {Evans}}]{Gibbons2014}
{Gibbons}, S.~L.~J., {Belokurov}, V., \& {Evans}, N.~W. 2014, \mnras, 445, 3788, \dodoi{10.1093/mnras/stu1986}

\bibitem[{{G{\'o}mez} {et~al.}(2015){G{\'o}mez}, {Besla}, {Carpintero}, {Villalobos}, {O'Shea}, \& {Bell}}]{Gomez2015}
{G{\'o}mez}, F.~A., {Besla}, G., {Carpintero}, D.~D., {et~al.} 2015, \apj, 802, 128, \dodoi{10.1088/0004-637X/802/2/128}

\bibitem[{{Hernquist}(1990)}]{Hernquist1990}
{Hernquist}, L. 1990, \apj, 356, 359, \dodoi{10.1086/168845}

\bibitem[{{Huang} \& {Carlberg}(1997)}]{Huang1997}
{Huang}, S., \& {Carlberg}, R.~G. 1997, \apj, 480, 503, \dodoi{10.1086/303977}

\bibitem[{{Hunter} \& {Toomre}(1969)}]{Hunter1969}
{Hunter}, C., \& {Toomre}, A. 1969, \apj, 155, 747, \dodoi{10.1086/149908}

\bibitem[{{Hwang} {et~al.}(2013){Hwang}, {Park}, \& {Choi}}]{Hwang2013}
{Hwang}, J.-S., {Park}, C., \& {Choi}, J.-H. 2013, Journal of Korean Astronomical Society, 46, 1, \dodoi{10.5303/JKAS.2013.46.1.1}

\bibitem[{{Kallivayalil} {et~al.}(2013){Kallivayalil}, {van der Marel}, {Besla}, {Anderson}, \& {Alcock}}]{Kallivayalil2013}
{Kallivayalil}, N., {van der Marel}, R.~P., {Besla}, G., {Anderson}, J., \& {Alcock}, C. 2013, \apj, 764, 161, \dodoi{10.1088/0004-637X/764/2/161}

\bibitem[{{Kerr}(1957)}]{Kerr1957}
{Kerr}, F.~J. 1957, \aj, 62, 93, \dodoi{10.1086/107466}

\bibitem[{{Koposov} {et~al.}(2019){Koposov}, {Belokurov}, {Li}, {Mateu}, {Erkal}, {Grillmair}, {Hendel}, {Price-Whelan}, {Laporte}, {Hawkins}, {Sohn}, {del Pino}, {Evans}, {Slater}, {Kallivayalil}, {Navarro}, \& {Orphan Aspen Treasury Collaboration}}]{Koposov2019}
{Koposov}, S.~E., {Belokurov}, V., {Li}, T.~S., {et~al.} 2019, \mnras, 485, 4726, \dodoi{10.1093/mnras/stz457}

\bibitem[{{Koposov} {et~al.}(2023){Koposov}, {Erkal}, {Li}, {Da Costa}, {Cullinane}, {Ji}, {Kuehn}, {Lewis}, {Pace}, {Shipp}, {Zucker}, {Bland-Hawthorn}, {Lilleengen}, {Martell}, \& {S5 Collaboration}}]{Koposov2023}
{Koposov}, S.~E., {Erkal}, D., {Li}, T.~S., {et~al.} 2023, \mnras, 521, 4936, \dodoi{10.1093/mnras/stad551}

\bibitem[{Kullback \& Leibler(1951)}]{Kullback-Leibler1951}
Kullback, S., \& Leibler, R.~A. 1951, The Annals of Mathematical Statistics, 22, 79 , \dodoi{10.1214/aoms/1177729694}

\bibitem[{{Laporte} {et~al.}(2018){Laporte}, {G{\'o}mez}, {Besla}, {Johnston}, \& {Garavito-Camargo}}]{Laporte2018}
{Laporte}, C. F.~P., {G{\'o}mez}, F.~A., {Besla}, G., {Johnston}, K.~V., \& {Garavito-Camargo}, N. 2018, \mnras, 473, 1218, \dodoi{10.1093/mnras/stx2146}

\bibitem[{{Law} \& {Majewski}(2010)}]{Law2010}
{Law}, D.~R., \& {Majewski}, S.~R. 2010, \apj, 714, 229, \dodoi{10.1088/0004-637X/714/1/229}

\bibitem[{{Li} {et~al.}(2023){Li}, {Huang}, {Liu}, {Beers}, \& {Zhang}}]{Li2023}
{Li}, X.-Y., {Huang}, Y., {Liu}, G.-C., {Beers}, T.~C., \& {Zhang}, H.-W. 2023, \apj, 944, 88, \dodoi{10.3847/1538-4357/acadd5}

\bibitem[{{Lilleengen} {et~al.}(2023){Lilleengen}, {Petersen}, {Erkal}, {Pe{\~n}arrubia}, {Koposov}, {Li}, {Cullinane}, {Ji}, {Kuehn}, {Lewis}, {Mackey}, {Pace}, {Shipp}, {Zucker}, {Bland-Hawthorn}, {Hilmi}, \& {S5 Collaboration}}]{Lilleengen2023}
{Lilleengen}, S., {Petersen}, M.~S., {Erkal}, D., {et~al.} 2023, \mnras, 518, 774, \dodoi{10.1093/mnras/stac3108}

\bibitem[{{Luri} {et~al.}(2018){Luri}, {Brown}, {Sarro}, {Arenou}, {Bailer-Jones}, {Castro-Ginard}, {de Bruijne}, {Prusti}, {Babusiaux}, \& {Delgado}}]{Luri2018}
{Luri}, X., {Brown}, A.~G.~A., {Sarro}, L.~M., {et~al.} 2018, \aap, 616, A9, \dodoi{10.1051/0004-6361/201832964}

\bibitem[{{Mendelsohn} {et~al.}(2022){Mendelsohn}, {Newberg}, {Shelton}, {Widrow}, {Thompson}, \& {Grillmair}}]{Mendelsohn2022}
{Mendelsohn}, E.~J., {Newberg}, H.~J., {Shelton}, S., {et~al.} 2022, \apj, 926, 106, \dodoi{10.3847/1538-4357/ac498a}

\bibitem[{{Navarro} {et~al.}(1997){Navarro}, {Frenk}, \& {White}}]{Navarro1997}
{Navarro}, J.~F., {Frenk}, C.~S., \& {White}, S. D.~M. 1997, \apj, 490, 493, \dodoi{10.1086/304888}

\bibitem[{{Nibauer} {et~al.}(2024){Nibauer}, {Bonaca}, {Lisanti}, {Erkal}, \& {Hastings}}]{Nibauer2024}
{Nibauer}, J., {Bonaca}, A., {Lisanti}, M., {Erkal}, D., \& {Hastings}, Z. 2024, \apj, 969, 55, \dodoi{10.3847/1538-4357/ad4299}

\bibitem[{{Patel} {et~al.}(2020){Patel}, {Kallivayalil}, {Garavito-Camargo}, {Besla}, {Weisz}, {van der Marel}, {Boylan-Kolchin}, {Pawlowski}, \& {G{\'o}mez}}]{Patel2020}
{Patel}, E., {Kallivayalil}, N., {Garavito-Camargo}, N., {et~al.} 2020, \apj, 893, 121, \dodoi{10.3847/1538-4357/ab7b75}

\bibitem[{{Pe{\~n}arrubia} {et~al.}(2016){Pe{\~n}arrubia}, {G{\'o}mez}, {Besla}, {Erkal}, \& {Ma}}]{Penarrubia2016}
{Pe{\~n}arrubia}, J., {G{\'o}mez}, F.~A., {Besla}, G., {Erkal}, D., \& {Ma}, Y.-Z. 2016, \mnras, 456, L54, \dodoi{10.1093/mnrasl/slv160}

\bibitem[{{Petersen} \& {Pe{\~n}arrubia}(2021)}]{Petersen2021}
{Petersen}, M.~S., \& {Pe{\~n}arrubia}, J. 2021, Nature Astronomy, 5, 251, \dodoi{10.1038/s41550-020-01254-3}

\bibitem[{{Pietrzy{\'n}ski} {et~al.}(2013){Pietrzy{\'n}ski}, {Graczyk}, {Gieren}, {Thompson}, {Pilecki}, {Udalski}, {Soszy{\'n}ski}, {Koz{\l}owski}, {Konorski}, {Suchomska}, {Bono}, {Moroni}, {Villanova}, {Nardetto}, {Bresolin}, {Kudritzki}, {Storm}, {Gallenne}, {Smolec}, {Minniti}, {Kubiak}, {Szyma{\'n}ski}, {Poleski}, {Wyrzykowski}, {Ulaczyk}, {Pietrukowicz}, {G{\'o}rski}, \& {Karczmarek}}]{Pietrzyski2013}
{Pietrzy{\'n}ski}, G., {Graczyk}, D., {Gieren}, W., {et~al.} 2013, \nat, 495, 76, \dodoi{10.1038/nature11878}

\bibitem[{{Plummer}(1911)}]{Plummer1911}
{Plummer}, H.~C. 1911, \mnras, 71, 460, \dodoi{10.1093/mnras/71.5.460}

\bibitem[{{Qian} {et~al.}(2022){Qian}, {Arshad}, \& {Bovy}}]{Qian2022}
{Qian}, Y., {Arshad}, Y., \& {Bovy}, J. 2022, \mnras, 511, 2339, \dodoi{10.1093/mnras/stac238}

\bibitem[{{Schommer} {et~al.}(1992){Schommer}, {Suntzeff}, {Olszewski}, \& {Harris}}]{Schommer1992}
{Schommer}, R.~A., {Suntzeff}, N.~B., {Olszewski}, E.~W., \& {Harris}, H.~C. 1992, \aj, 103, 447, \dodoi{10.1086/116074}

\bibitem[{{Shelton}(2018)}]{Shelton2018}
{Shelton}, S. 2018, PhD thesis, Rensselaer Polytechnic Institute, New York

\bibitem[{{Shelton} {et~al.}(2021){Shelton}, {Newberg}, {Weiss}, {Bauer}, {Arsenault}, {Widrow}, {Rayment}, {Desell}, {Judd}, {Magdon-Ismail}, {Mendelsohn}, {Newby}, {Rice}, {Szymanski}, {Thompson}, {Varela}, {Willett}, {Ulin}, \& {Newberg}}]{Shelton2021}
{Shelton}, S., {Newberg}, H.~J., {Weiss}, J., {et~al.} 2021, arXiv e-prints, arXiv:2102.07257, \dodoi{10.48550/arXiv.2102.07257}

\bibitem[{{Shipp} {et~al.}(2021){Shipp}, {Erkal}, {Drlica-Wagner}, {Li}, {Pace}, {Koposov}, {Cullinane}, {Da Costa}, {Ji}, {Kuehn}, {Lewis}, {Mackey}, {Simpson}, {Wan}, {Zucker}, {Bland-Hawthorn}, {Ferguson}, {Lilleengen}, \& {Lilleengen}}]{Shipp2021}
{Shipp}, N., {Erkal}, D., {Drlica-Wagner}, A., {et~al.} 2021, \apj, 923, 149, \dodoi{10.3847/1538-4357/ac2e93}

\bibitem[{{van der Marel} {et~al.}(2002){van der Marel}, {Alves}, {Hardy}, \& {Suntzeff}}]{Van2002}
{van der Marel}, R.~P., {Alves}, D.~R., {Hardy}, E., \& {Suntzeff}, N.~B. 2002, \aj, 124, 2639, \dodoi{10.1086/343775}

\bibitem[{{van der Marel} {et~al.}(2012){van der Marel}, {Besla}, {Cox}, {Sohn}, \& {Anderson}}]{van2012}
{van der Marel}, R.~P., {Besla}, G., {Cox}, T.~J., {Sohn}, S.~T., \& {Anderson}, J. 2012, \apj, 753, 9, \dodoi{10.1088/0004-637X/753/1/9}

\bibitem[{{van der Marel} \& {Kallivayalil}(2014)}]{Van2014}
{van der Marel}, R.~P., \& {Kallivayalil}, N. 2014, \apj, 781, 121, \dodoi{10.1088/0004-637X/781/2/121}

\bibitem[{{Vasiliev} {et~al.}(2021){Vasiliev}, {Belokurov}, \& {Erkal}}]{Vasiliev2021}
{Vasiliev}, E., {Belokurov}, V., \& {Erkal}, D. 2021, \mnras, 501, 2279, \dodoi{10.1093/mnras/staa3673}

\bibitem[{{Vera-Ciro} \& {Helmi}(2013)}]{Vera2013}
{Vera-Ciro}, C., \& {Helmi}, A. 2013, \apjl, 773, L4, \dodoi{10.1088/2041-8205/773/1/L4}

\bibitem[{{Verlet}(1967)}]{Verlet1967}
{Verlet}, L. 1967, Physical Review, 159, 98, \dodoi{10.1103/PhysRev.159.98}

\bibitem[{{Wang} {et~al.}(2022){Wang}, {Hammer}, \& {Yang}}]{Wang2022}
{Wang}, J., {Hammer}, F., \& {Yang}, Y. 2022, \mnras, 515, 940, \dodoi{10.1093/mnras/stac1640}

\bibitem[{{Wang} {et~al.}(2019){Wang}, {Hammer}, {Yang}, {Ripepi}, {Cioni}, {Puech}, \& {Flores}}]{Wang2019}
{Wang}, J., {Hammer}, F., {Yang}, Y., {et~al.} 2019, \mnras, 486, 5907, \dodoi{10.1093/mnras/stz1274}

\bibitem[{{Watkins} {et~al.}(2024){Watkins}, {van der Marel}, \& {Bennet}}]{Watkins2024}
{Watkins}, L.~L., {van der Marel}, R.~P., \& {Bennet}, P. 2024, \apj, 963, 84, \dodoi{10.3847/1538-4357/ad1f58}

\end{thebibliography}

\end{document}